\documentclass[aps,prx,twocolumn,superscriptaddress,showpacs]{revtex4-1}

\usepackage{amsmath}
\usepackage{amsfonts}
\usepackage{amssymb}
\usepackage{graphicx}
\usepackage{color}
\usepackage{bm}
\usepackage{hyperref}
\usepackage{layouts}

\begin{document}

\title{Exponential Lifetime Improvement in Topological Quantum Memories}

\author{Charles-Edouard Bardyn}
\affiliation{Institute for Quantum Information and Matter, Caltech, Pasadena, California 91125, USA}
\author{Torsten Karzig}
\affiliation{Station Q, Microsoft Research, Santa Barbara, California 93106-6105, USA}
\affiliation{Institute for Quantum Information and Matter, Caltech, Pasadena, California 91125, USA}

\begin{abstract}
We propose a simple yet efficient mechanism for passive error correction in topological quantum memories. Our scheme relies on driven-dissipative ancilla systems which couple to local excitations (anyons) and make them ``sink'' in energy, with no required interaction among ancillae or anyons. Through this process, anyons created by some thermal environment end up trapped in potential ``trenches'' that they themselves generate, which can be interpreted as a ``memory foam'' for anyons. This self-trapping mechanism provides an energy barrier for anyon propagation, and removes entropy from the memory by favoring anyon recombination over anyon separation (responsible for memory errors). We demonstrate that our scheme leads to an exponential increase of the memory-coherence time with system size $L$, up to an upper bound $L_\mathrm{max}$ which can increase exponentially with $\Delta/T$, where $T$ is the temperature and $\Delta$ is some energy scale defined by potential trenches. This results in a double exponential increase of the memory time with $\Delta/T$, which greatly improves over the Arrhenius (single-exponential) scaling found in typical quantum memories. 
\end{abstract}

\maketitle


\section{Introduction}

The ease with which classical information can be stored is often taken for granted. Yet achieving the analog of simple tasks such as recording a list of bits on a long-lasting piece of paper remains extremely challenging in the quantum realm. Despite  tremendous progress in manipulating individual quantum ``objects'' such as electrons or photons, \emph{self-correcting} classical memories --- memories that can store bits at finite temperature for arbitrarily long periods of time without active error correction~\cite{Dennis2002,Bacon2006,Bravyi2009,Yoshida2011} --- currently have no known quantum counterpart. The only exceptions are theoretical models requiring more than three spatial dimensions~\cite{Dennis2002,Alicki2010}.

Quantum bits (qubits) require more protection than classical bits as they can be in a coherent superposition of two states, which is particularly prone to dephasing due to random energy fluctuations induced by a thermal environment. To suppress this source of decoherence, Kitaev and Preskill introduced the concept of \emph{topological} quantum memory~\cite{Dennis2002} where quantum information is stored in the ground-state subspace of a Hamiltonian with topological order~\cite{Wen1990}. Topology guarantees that states belonging to this subspace remain degenerate under (weak and static) local perturbations, in the limit of a large system~\cite{Bravyi2010,Michalakis2013}.
Quantum information is encoded in a non-local way, which shields it from the local perturbations induced by typical thermal environments.

\begin{figure}[!htb]
    \begin{center}
        \includegraphics[width=0.92\linewidth]{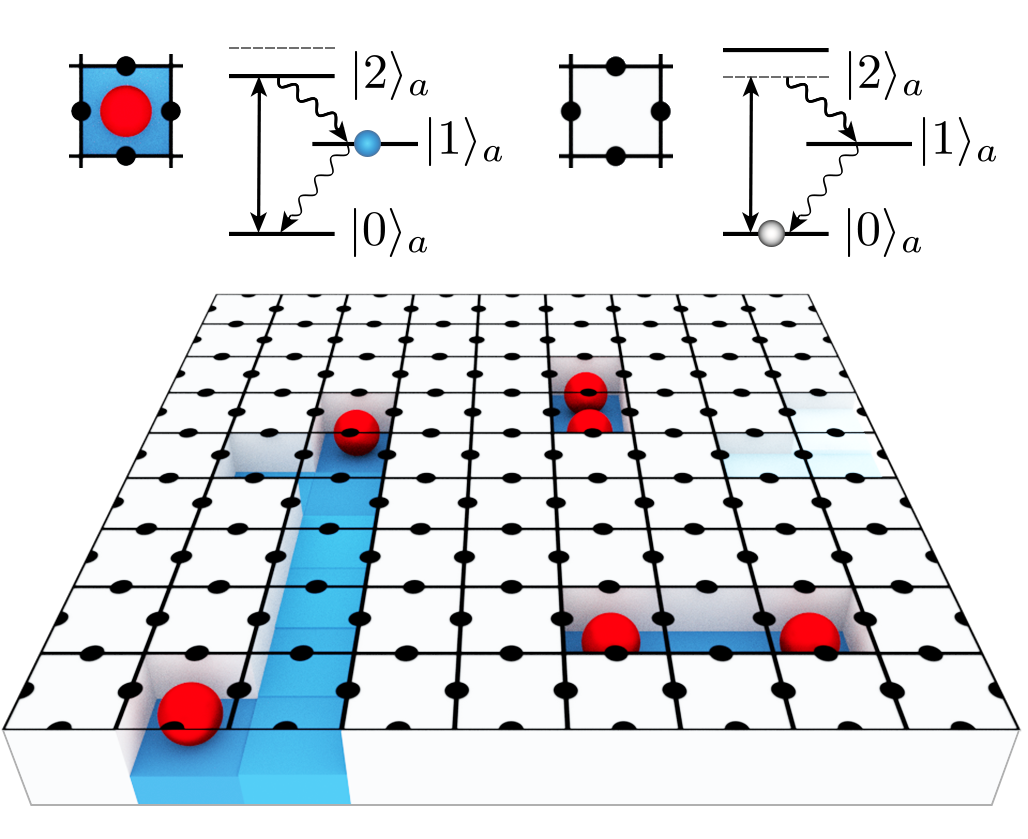}
        \caption{Efficient passive error correction provided by a ``memory foam'' for anyons (applied, here, to the toric code). Each plaquette of four spins (black dots) on the toric-code lattice is coupled to a driven-dissipative ancilla system whose purpose is to lower the plaquette energy (indicated by height and white-to-blue color scale) whenever the plaquette is visited by an anyon (in red). The ancilla can be regarded as a quantum three-level system with a coherent drive (black double arrow) coupling level $| 0 \rangle_a$ to level $| 2 \rangle_a$, with fast subsequent decay to some metastable level $| 1 \rangle_a$ and slower decay back to $| 0 \rangle_a$ (we describe an alternative practical realization based on cavity-QED systems in Appendix~\ref{app:ancillaSystems}). When a plaquette becomes occupied (top left), level $| 2 \rangle_a$ falls into resonance with the drive and level $| 1 \rangle_a$ is quickly populated (blue dot). In contrast, when no anyon is present (top right), level $| 2 \rangle_a$ is off-resonant and any occupation of level $| 1 \rangle_a$ slowly decays back to level $| 0 \rangle_a$ (grey dot). Crucially, the population of level $| 1 \rangle_a$ acts back on the toric-code memory by lowering the local plaquette energy. As a result, anyons (created in pairs) become trapped in potential ``trenches'' that they self-generate (in blue). This ``memory-foam'' effect makes it much more likely for anyons to recombine than to separate and cause memory errors.}
        \label{fig:setup}
    \end{center}
\end{figure}

A paradigmatic example of topological quantum memory --- the 2D \emph{toric code} --- was introduced by Kitaev in a pioneering work~\cite{Kitaev2003}. Subsequent studies demonstrated, however, that Kitaev's toric code does not provide passive protection against errors induced by a thermal environment~\cite{Alicki2009}. The crux of the issue is that local excitations created by thermal fluctuations --- known as \emph{anyons} --- are essentially free to diffuse over large distances with no energy cost, which eventually leads to harmful non-local perturbations. The possibility of self-correction was ruled out in broad classes of 2D and 3D models for similar reasons~\cite{Kay2008,Bravyi2009,Pastawski2010,Yoshida2011,Landon-Cardinal2013,Pastawski2015}. Despite these no-go theorems, several strategies have been developed to passively prolong the lifetime of topological quantum memories: long-range interactions between anyons~\cite{Hamma2009,Chesi2010,Pedrocchi2013,Hutter2014}, energy~\cite{Bacon2006,Haah2011,Bravyi2011,Bravyi2013,Kim2012,Michnicki2012,Michnicki2014} and entropic barriers~\cite{Brown2014} to suppress anyon propagation, disorder to localize the anyons~\cite{Stark2011,Wootton2011,Bravyi2012}, and engineered dissipation to remove entropy and excitations from the system~\cite{Pastawski2011,Fujii2014,Kapit2015} (see Refs.~\cite{Brown2014_2,Landon-Cardinal2015,Terhal2015} for recent reviews). Classical cellular automata effectively mediating long-range interactions between anyons have also recently been proposed to actively correct errors~\cite{Herold2015,Herold2015_2}.

In this work, we propose an efficient mechanism to passively prolong the lifetime of topological quantum memories based on stabilizer codes~\cite{Gottesman1997}, focusing on Kitaev's 2D toric code as an example. Our scheme relies on the introduction of driven-dissipative quantum systems (ancillae) which couple to the memory locally. When anyons are created by local perturbations, the energy of the corresponding stabilizers (or ``plaquettes'') is quickly reduced due to the memory-ancillae coupling, which effectively makes the anyons ``sink'' in energy. When an anyon moves to a different plaquette, the energy of the previous one remains lower for a certain time, such that anyons become trapped in an extended potential well (or ``trench'') that they generate (see Fig.~\ref{fig:setup}). Since escaping a trench requires to overcome an energy barrier, anyons tend to recombine instead of separating by large distances, which efficiently suppresses memory errors. This powerful dissipative error-correction mechanism can be regarded as a ``memory foam'' for anyons.

To demonstrate the efficiency of our scheme, we investigate the dynamics of anyons in a standard paradigmatic model: Kitaev's 2D toric code coupled to a bosonic thermal environment (Ohmic bath)~\cite{Leggett1987}. Remarkably, we find that the coupling to driven-dissipative ancilla systems induces a significant free-energy barrier to errors: for a constant ancilla-memory coupling strength $U$, the memory lifetime increases exponentially with system size $L$, up to an upper bound independent of $L$. Most importantly, the maximum $\tau_\mathrm{max}$ can increase \emph{double} exponentially as the bath temperature is lowered, i.e., $\tau_\mathrm{max} \sim \exp(c_1 \mathrm{e}^{\beta c_2})$, where $c_1$ and $c_2$ are positive constants and $\beta = 1/T$ is the inverse bath temperature (we set the Boltzmann constant to unity). This behavior, which holds in the low-temperature regime $T \ll U$, is in stark contrast to the scaling $\tau_\mathrm{max} \sim \exp(\beta\Delta)$ naively expected from the Arrhenius law for quantum memories with a gap $\Delta$. It also strikingly differs from the super-Arrhenius scaling $\tau_\mathrm{max} \sim \exp(c \beta^2)$ (with $c > 0$) obtained in recent proposals for passive error correction in topological quantum memories~\cite{Haah2011,Castelnovo2011,Brown2014}.

The coherence time of quantum memories is generally governed by the interplay between energy barriers and entropic contributions to errors~\cite{Brown2014_2}. In Kitaev's 2D toric code, though the creation of anyons requires an energy of the order of the gap, the energy barrier that must be overcome to separate anyons by a distance $\ell$ and create an error is independent of $\ell$. Anyons are free to diffuse, and entropic effects play a prominent role since the number of pathways leading to an error increases with system size~\cite{Brown2014_2}. Remarkably, the situation changes drastically upon introducing driven-dissipative ancilla systems as per our proposal: When anyons continuously sink into potential wells, the energy required for them to diffuse a distance $\ell$ away from each other increases (linearly) with $\ell$. In addition to this large energy barrier, entropic contributions are also strongly suppressed. Indeed, the finite lifetime of potential wells at location previously visited by anyons generates preferred pathways for anyon recombination, thereby suppressing memory errors.

A number of theoretical proposals have led to memory lifetimes that scale polynomially with system size, up to an upper bound that depends on temperature with super-Arrhenius scaling of the form $\tau_\mathrm{max} \sim \exp(c \beta^2)$ (with $c > 0$)~\cite{Haah2011,Castelnovo2011}. This behavior is commonly referred to as partial self-correction, as opposed to true self-correction characterized by a coherence time that increases without bound with system size~\cite{Brown2014_2}. It is generally attributed to the existence of an energy barrier that grows logarithmically with the separation between local excitations, which is the most favorable scaling achievable in a wide family of quantum memories based on spin models with translation invariance~\cite{Haah2013}. Recent studies have explored ways to surpass these results by breaking translation invariance~\cite{Michnicki2012,Michnicki2014,Brell2014}. So far, however, the complexity of these models has made it difficult to investigate their self-correction properties (see, e.g.,~\cite{Brown2014_2}). The scheme that we propose in this work lends itself more easily to theoretical analysis. As opposed to known models for partial self-correction~\cite{Haah2011,Castelnovo2011}, it does not require to break translation invariance and does not introduce a ground-state degeneracy that depends on system size. Most importantly, it leads to an exponential improvement of the memory lifetime with increasing system size, which, even if bounded, is reminiscent of true self-correcting models such as the 4D toric code~\cite{Dennis2002}.

\subsection{Structure of the paper}

This paper is organized as follows: First, in Sec.~\ref{sec:model}, we present our model. We start by briefly reviewing Kitaev's 2D toric code which will serve as a ``toy-model'' topological quantum memory to illustrate our scheme. We then introduce the driven-dissipative ancillae that lie at the heart of our proposal and consider a standard error model based on a generic type of (bosonic) thermal bath. Next, in Sec.~\ref{sec:singleError}, we examine the low-temperature regime in which a \emph{single} pair of anyons is present, with no additional pair creation. To gain intuition about the dynamics of a single pair, we consider a simple toy model where anyons generate a self-trapping potential that (i) develops instantaneously on plaquettes that they visit, (ii) does not decay, and (iii) can only expand along one dimension (forming a 1D potential trench). We then argue that relevant potential trenches are indeed expected to be (mostly) one-dimensional, and study both analytically and numerically the effects of relaxing assumptions (i) and (ii). In Sec.~\ref{sec:memoryFiguresOfMerit}, we present the figures of merit of the quantum memory obtained through our scheme. We discuss the effects arising in the general case where multiple anyon pairs are present, and provide qualitative estimates for the memory lifetime. We support our claims by extensive numerical computations. Finally, in Sec.~\ref{sec:conclusion}, we summarize our results and discuss their implications. We then provide additional details in Appendices.

\section{The model}
\label{sec:model}

The scheme that we propose in this work is directly relevant to any kind of topological quantum memory based on stabilizer codes~\cite{Gottesman1997}. To provide a concrete discussion and quantitative results, we focus on Kitaev's paradigmatic (2D) toric code. We first briefly recall the details of this model.

\subsection{Kitaev's toric code}

Kitaev's toric code is defined on a two-dimensional surface with periodic boundary conditions (a torus). It consists of quantum two-level systems (``spins'') arranged on a square lattice, as depicted in Fig.~\ref{fig:setup}. These ``physical'' qubits are used to encode two ``logical'' qubits in a subspace identified by mutually commuting quasi-local operators (or ``stabilizers'') of the form
\begin{equation}
    A_s = \prod_{j \in s} \sigma_j^x, \quad B_p = \prod_{j \in p} \sigma_j^z,
    \label{eq:stabilizers}
\end{equation}
where $\sigma_j^x$ and $\sigma_j^z$ are Pauli matrices and $j \in s (p)$ denotes the set of four spins $j$ belonging to a ``star'' (around vertices) or ``plaquette'' (around elementary squares; see Fig.~\ref{fig:setup}), respectively. The subspace where logical qubits are stored is spanned by the eigenstates $|\psi\rangle$ of star and plaquette operators ($A_s$ and $B_p$) with eigenvalue $+1$, i.e., such that $A_s |\psi\rangle = B_p |\psi\rangle = |\psi\rangle$. This ``stabilizer space'' coincides with the ground-state subspace of the toric-code Hamiltonian
\begin{equation}
    H_\mathrm{TC} = -\frac{J_s}{2} \sum_s A_s -\frac{J_p}{2} \sum_p B_p,
    \label{eq:toricCodeHam}
\end{equation}
where $J_s$ and $J_p$ are positive constants which we refer to as ``star'' and ``plaquette'' energies, respectively. In this setting, stars and plaquettes can be regarded as occupied by a quasiparticle excitation whenever the system lies in a eigenstate of the corresponding operator ($A_s$ or $B_p$) with eigenvalue $-1$. These excitations are (Abelian) anyons~\cite{Kitaev2003}, and the stabilizer space (or ground-state subspace) can be seen as the associated vacuum.

The lifetime of the quantum information stored in the toric code crucially hinges on the dynamics of anyons. In general, local ``errors'' (defined by local Pauli operators) create, move, or annihilate anyons. Starting from the anyonic vacuum, for example, a ``bit-flip'' error $\sigma_j^x$ acting on a physical qubit $j$ creates a pair of anyons on neighboring plaquettes (since $\sigma_j^x$ anticommutes with two plaquette operators). If an anyon is already present on one of these plaquettes, the bit flip moves it to the neighboring one, and if a pair of anyons is already occupying the neighboring plaquettes, the bit flip annihilates it. The process of creating an anyon pair, moving the anyons, and recombining them requires a product $\prod_{j \in \mathcal{C}} \sigma_j^x$ of local errors along a closed loop $\mathcal{C}$ on the lattice. Although it is clear that such loops leave the system in its ground-state or stabilizer space, loops that ``go around'' the torus (such that they cannot be contracted) do modify the state of the logical qubits. Indeed, the Pauli operators associated with these qubits correspond to products of local spin operators $\sigma_j^{x,z}$ along loops winding around the torus (two in each direction)~\cite{Kitaev2003}. These loops can be defined arbitrarily provided that they wind around the torus and that the resulting operators commute with all plaquette and star operators. The quantum information encoded in the toric code is then robust against local errors provided that topologically non-trivial loops are not created. We now present an efficient scheme to suppress logical errors in a passive way, with no need to constantly monitor the anyon positions and actively recombine anyons coming from the same pair~\footnote{As discussed in Sec.~\ref{sec:conclusion}, our scheme could also be used for active error correction or decoding.}.

\subsection{Ancilla systems}
\label{sec:ancillaSystems}

The creation of anyons (in pairs) on stars or plaquettes is suppressed by an energy gap $2J_s$ or $2J_p$, respectively [see Eq.~\eqref{eq:toricCodeHam}]. Once anyons are created, however, no additional energy cost is required for them to diffuse around the torus and cause a logical error. To generate an energy cost and hinder such diffusion, we introduce on each plaquette and star an ancilla system which performs a simple local task: It repopulates a slowly-decaying (metastable) state whenever an anyon visits the plaquette (star), and this population acts back on the system by effectively lowering the plaquette (star) energy $J_p$ ($J_s$) [see Fig.~\ref{fig:setup}]. Below we focus on plaquettes, for simplicity, restricting ourselves to errors of the type $\sigma_j^x$ (bit flips). Our scheme can be readily extended to stars, in which case protection against all types of local errors would be provided.

Our proposal consists in complementing each plaquette $p$ by an ancilla system $a_p$ that acts on the original system according to the effective Hamiltonian
\begin{equation}
    H_{p, \mathrm{eff}}(t) = \frac{n_{a_p}(t)U}{2} B_p,
    \label{eq:effectiveHamAncilla}
\end{equation}
where $U > 0$ is a coupling constant and $n_{a_p}(t) \leq 1$ is the population of the metastable ancilla state at time $t$. Remembering the form of the toric-code Hamiltonian~\eqref{eq:toricCodeHam}, one sees that $H_{p, \mathrm{eff}}$ effectively shifts the energy of the associated plaquette from $J_p$ to $J_p - n_{a_p}U$ (we assume that $U < J_p$ so that this energy remains positive). Since this corresponds to the energy of an anyon on the plaquette, the energy shift $-n_{a_p}U$ can be interpreted as a \emph{local effective potential well} for anyons. The coupling constant $U$ corresponds to the maximum depth of this potential.

The time evolution of the ancilla state population $n_{a_p}$ (or potential depth $n_{a_p}U$) plays a key role in our scheme. We assume that it decays at a slow rate $\gamma_\mathrm{dec}$ (corresponding to the lifetime of the metastable ancilla state), and, most importantly, that it is ``repumped'' to unity at a fast rate $\gamma_\mathrm{pump} \gg \gamma_\mathrm{dec}$ whenever an anyon is present on the plaquette. We present in Appendix~\ref{app:ancillaSystems} more details regarding the implementation of an ancilla system that would provide such dynamics and interact with the toric code according to Eq.~\eqref{eq:effectiveHamAncilla}. Due to the ancilla-pumping mechanism, anyons continuously ``sink'' in energy (over a time $\sim 1/\gamma_\mathrm{pump}$ and by a maximum amount of $-U$). An effective potential well of depth $n_{a_p}U$ develops on each plaquette that they occupy, and this potential decays on a slow time scale $\sim 1/\gamma_\mathrm{dec}$ as they leave the latter.

The ancilla systems introduced above provide a simple mechanism for error correction:  When an anyon pair is created, both anyons quickly sink into potential wells of depth $\sim U$, thereby opening up a potential well of area $2$ (in units of plaquette area). Trapped in this potential, the probability that they diffuse away from each other instead of recombining can be greatly reduced, as we demonstrate below. Every time they manage to separate further (by paying an energy cost $\sim U$), they extend the potential well in which they are trapped, which makes it more likely for them to retrace their steps and recombine. This mechanism lies at the heart of our proposal. It generates an anyon dynamics that prolongs the lifetime of the memory. The fact that potential wells decay over a slow time scale $\sim 1/\gamma_\mathrm{dec}$ further complicates the dynamics, but is necessary to avoid the proliferation of potential wells everywhere in the system (in which case anyons would diffuse freely as in the original toric code).

We remark that the ancilla systems record information about the toric code and act back on the latter in a quantum non-demolition way, leaving the system in an eigenstate of plaquette and star operators. Indeed, $H_{p, \mathrm{eff}}$ [Eq.~\eqref{eq:effectiveHamAncilla}] commutes with the toric-code Hamiltonian [Eq.~\eqref{eq:toricCodeHam}]. Our scheme is robust against small local perturbations of the toric-code-ancilla interaction $H_{p, \mathrm{eff}}$, since the ground-state degeneracy of the toric code is well known to be stable against generic perturbations that are local and small as compared to the energy gap $2J$ (we set $J_s = J_p \equiv J$ in Eq.~\eqref{eq:toricCodeHam}, for simplicity)~\cite{Kitaev2003}. The more relevant issue concerns the robustness of the system as a quantum memory \emph{at finite temperature}. To address this issue, we now introduce a model thermal bath that weakly couples to the system.

\subsection{Thermal bath}

To model the interaction of the system with a typical thermal environment, we introduce a bosonic bath of harmonic oscillators and assume that each physical spin of the system is weakly coupled to this bath as described by the paradigmatic spin-boson model of quantum dissipation~\cite{Leggett1987}. More specifically, we consider a local spin-bath coupling of the form $\sim \sigma_j^x \sum_i \lambda_i (a^\dagger_i + a_i)$ (inducing bit-flip errors), where $\lambda_i$ is the coupling amplitude to a particular bosonic mode $i$ with creation (annihilation) operator $a^\dagger_i$ ($a_i$). After a standard master-equation treatment of the system-bath interaction (see, e.g., Ref.~\cite{Breuer2007}), one finds a rate equation for the toric-code dynamics
\begin{equation}
    \dot{p}_m = \sum_{n} \left[ \gamma(\omega_{mn}) \, p_n - \gamma(\omega_{nm}) \, p_m \right],
    \label{eq:rateEquation}
\end{equation}
where $p_m$ is the probability that the toric-code system is in state $|\psi_m\rangle$ and $\gamma(\omega_{nm})$ is the transition rate to state $|\psi_n\rangle$, which depends on the energy difference $\omega_{nm} \equiv E_n - E_m$ between initial and final states. The explicit form of the rates derived from the spin-boson model is
\begin{equation}
    \gamma(\omega) = \kappa_\lambda \left| \frac{\omega^\lambda}{e^{\omega/T} - 1} \right| e^{-|\omega|/\omega_c},
    \label{eq:transitionRate}
\end{equation}
where $\kappa_\lambda > 0$ is a coupling constant, $T \equiv 1/\beta$ is the bath temperature, and $\omega_c$ is an energy cutoff which we assume to be much larger than all relevant energy scales, for simplicity, such that $e^{-|\omega|/\omega_c} \approx 1$. The constant $\lambda$ characterises the low-energy behavior of $\gamma(\omega)$. Although our scheme applies more broadly, we choose $\lambda = 1$ and define $\kappa \equiv \kappa_1$. This corresponds to a very common type of bath usually referred to as ``Ohmic''~\cite{Leggett1987}. Irrespective of $\lambda$, Eq.~\eqref{eq:transitionRate} ensures that the rates satisfy the detailed balance condition $\gamma(-\omega) = e^{-\omega/T}\gamma(\omega)$, which would lead to a thermal (Gibbs) steady state in the absence of ancilla systems. We note that the generic bath considered here induces uncorrelated bit-flip errors, as would typically be the case in systems with (quasi-)local interactions.

\section{Correction of a single error}
\label{sec:singleError}

In the toric code, memory errors occur due to the creation and subsequent separation of anyon pairs. For a finite density $\rho$ of anyons, an error occurs when anyons separate over distances $\sim 1/\sqrt{\rho}$ corresponding to the typical distance between anyon pairs. In this section, we start by examining the dynamics of a \emph{single} anyon pair in the presence of ancilla systems as introduced above. We estimate, in particular, the probability $P_\mathrm{sep}(\ell)$ that an anyon pair separates by a distance $\ell$ before recombining. For a well-defined potential trench of depth $U \gg T$, this probability is governed by the small parameter $\alpha_\uparrow \equiv \gamma(U)/\gamma(0)$ which determines the probability of escaping the trench instead of simply diffusing inside it~\footnote{In particular, the probability that an isolated anyon located at the end of a 1D trench ``jumps out'' of the latter (thereby extending it) is equal to $\gamma(U)/[3\gamma(U) + \gamma(0)] = 1/(3\alpha + 1)$.}. 
For small enough $\alpha_\uparrow \ll 1$, we find that $P_\mathrm{sep}(\ell) \propto (\alpha_\uparrow)^\ell$. This exponential decay lies at the heart of our scheme.

\subsection{Instructive toy model: single anyon pair in a 1D potential trench}
\label{sec:1DToyModel}

The interplay of intra-trench anyon diffusion and occasional trench expansion makes it challenging to provide a general analytical description of the separation probability $P_\mathrm{sep}(\ell)$. Difficulties arise, in particular, from the fact that at least two anyons diffuse simultaneously within a trench. To gain insight into this dynamics, we first examine an instructive toy model based on the following simplifications: (i) We consider a \emph{single} pair of anyons. (ii) We assume that potential wells develop instantaneously on plaquettes where anyons are located (leading to the formation of a trench), and that local potentials do not decay over time (i.e., we consider an infinite pump rate and a vanishing decay rate; see Sec.~\ref{sec:ancillaSystems}). (iii) We assume that the resulting potential trench is purely one-dimensional. Indeed, 1D trenches minimize the number of trench extensions and thus provide the most significant contribution to $P_\mathrm{sep}(\ell)$ when extensions are very costly. As we show in Appendix~\ref{app:applicability1D}, we expect 1D trenches to dominate for length scales $\ell < \ell_{1\mathrm{D}}$ with
\begin{equation}
    \ell_{1\mathrm{D}} \approx \left( \frac{9}{2 \alpha_\uparrow} \right)^{1/3}.
    \label{eq:ell1D}
\end{equation}
Finally (iv), we assume that one of the anyons is pinned at its initial position (i.e., at one end of the trench). This assumption allows us to treat the problem as a ``single-particle problem'', and yields an upper bound $P^{(0)}_\mathrm{sep}(\ell)$ for the pair-separation probability. Indeed, pinning one anyon at one end of the trench minimizes the recombination probability by maximizing the distance between the two anyons. Note that an upper bound on the separation probability leads to a lower bound on the memory-coherence time (see Sec.~\ref{sec:memoryFiguresOfMerit}).

We consider the generic initial configuration where a pair of anyons is created by a local perturbation, such that both anyons quickly (here, instantaneously) end up trapped in a 1D potential trench of length $2$ (in units of plaquette length). We define the anyon positions as $x = 0$ and $x = 1$, respectively, and assume that the left anyon (at $x = 0$) is pinned. The separation probability $P^{(0)}_\mathrm{sep}(\ell)$ is then defined as the probability that the right anyon reaches a maximum separation $x = \ell$ before recombining with the left one. This probability can be constructed recursively. Assuming that the right anyon just extended the trench to reach $x = \ell-1$, two events can eventually occur: Either the anyon further extends the trench, or it recombines with its partner. The probability of a new extension can be expressed as
\begin{equation}
    \frac{P^{(0)}_\mathrm{sep}(\ell)}{P^{(0)}_\mathrm{sep}(\ell-1)} = \frac{\alpha_\uparrow}{1+\alpha_\uparrow} + \frac{1}{1+\alpha_\uparrow} \left[\bar{P}_\mathrm{rec}(\ell-2) \frac{P^{(0)}_\mathrm{sep}(\ell)}{P^{(0)}_\mathrm{sep}(\ell-1)} \right],
    \label{eq:Pl/Pl-1}
\end{equation}
which reflects two possibilities: Either the anyon ``jumps'' to the right ($x = \ell$) and, therefore, directly extends the trench (first term), or it jumps to the left ($x = \ell-2$) and eventually comes back to the starting point $x = \ell-1$ without recombining (second term), in which case recursion occurs. Here, $\bar{P}_\mathrm{rec}(\ell-2) \equiv 1 - P_\mathrm{rec}(\ell-2)$ denotes the probability that the anyon starting at $x = \ell-2$ comes back to $x = \ell-1$ without taking $\ell-2$ steps to the left and hence recombine. The probability of the complementary event (i.e., of not coming back to $x = \ell -1$ and instead recombining at $x = 0$) is mainly determined by the free 1D diffusion of the anyon in the potential trench, yielding $P_\mathrm{rec}(\ell-2) \sim 1/\ell$. The only difference stems from the fact that the last recombination step (when the anyon sits at $x = 1$ next to its partner) occurs at a different relative rate $\alpha_\downarrow \equiv \gamma(-[\Delta-2U])/\gamma(0) > 1$ as compared to free diffusion with rate $\gamma(0)$. As shown in Appendix~\ref{app:derivationPsep}, we find $P_\mathrm{rec}(\ell) = 1/(\ell+1/\alpha_\downarrow)$.

With the initial condition $P^{(0)}_\mathrm{sep}(1) = 1$, a straightforward iteration of Eq.~\eqref{eq:Pl/Pl-1} leads to the solution
\begin{equation}
    P^{(0)}_\mathrm{sep}(\ell) = \prod_{l=2}^{\ell} \frac{\alpha_\uparrow}{\alpha_\uparrow +\big( l-2+\alpha_\downarrow^{-1} \big)^{-1}}.
    \label{eq:1DResult}
\end{equation}
Since $0 < 1/\alpha_\downarrow < 1$, the contribution of each factor (each $l$) to Eq.~\eqref{eq:1DResult} is governed by the product $\alpha_\uparrow l$. This allows us to identify two regimes based on whether contributions with $\alpha_\uparrow l \ll 1$ or $\alpha_\uparrow l \gg 1$ dominate:

When $\alpha_\uparrow \ell \ll 1$ [such that $\alpha_\uparrow l \ll 1$ for all contributions in Eq.~\eqref{eq:1DResult}], the pair-separation probability decays to leading order exponentially with $\ell$, i.e., $P^{(0)}_\mathrm{sep}(\ell) \approx \alpha_\uparrow^\ell \, \ell! \sim (\alpha_\uparrow \ell/\mathrm{e})^\ell$. The $\ell !$ correction is a direct consequence of the fact that the probability to extend a trench of length $l$ increases linearly with $l$. Indeed, as the length increases, the right anyon must diffuse over a larger distance to be able to recombine with its partner. More specifically, the time required for the right anyon to reach the origin (and possibly recombine) when it just extended the trench is $t_{l} \sim l^{2}/\gamma(0)$. Since the anyon roughly spends a fraction $1/l$ of this time at the trench boundary where it can further extend the trench, the extension probability can be estimated as $\gamma(U) t_l/l \sim \alpha_\uparrow l$.

For $\alpha_\uparrow \ell \gg 1$ [i.e., for larger separations $\ell \gg 1/\alpha_\uparrow$ such that factors with $l \gg 1/\alpha_\uparrow$ appear in Eq.~\eqref{eq:1DResult}], the exponential suppression crosses over to a power-law decay $P^{(0)}_\mathrm{sep}(\ell) \sim (\alpha_\uparrow \ell)^{-1/\alpha_\uparrow}$. The value $P^{(0)}_\mathrm{sep}(1/\alpha_\uparrow) \sim \mathrm{e}^{-1/\alpha_\uparrow}$ reached at the crossover decreases double exponentially with decreasing temperature $T$, in the regime $U/T \gg 1$ of interest where $1/\alpha_\uparrow \sim \mathrm{e}^{U/T}$. 

\begin{figure}[t]
    \includegraphics[width=\columnwidth]{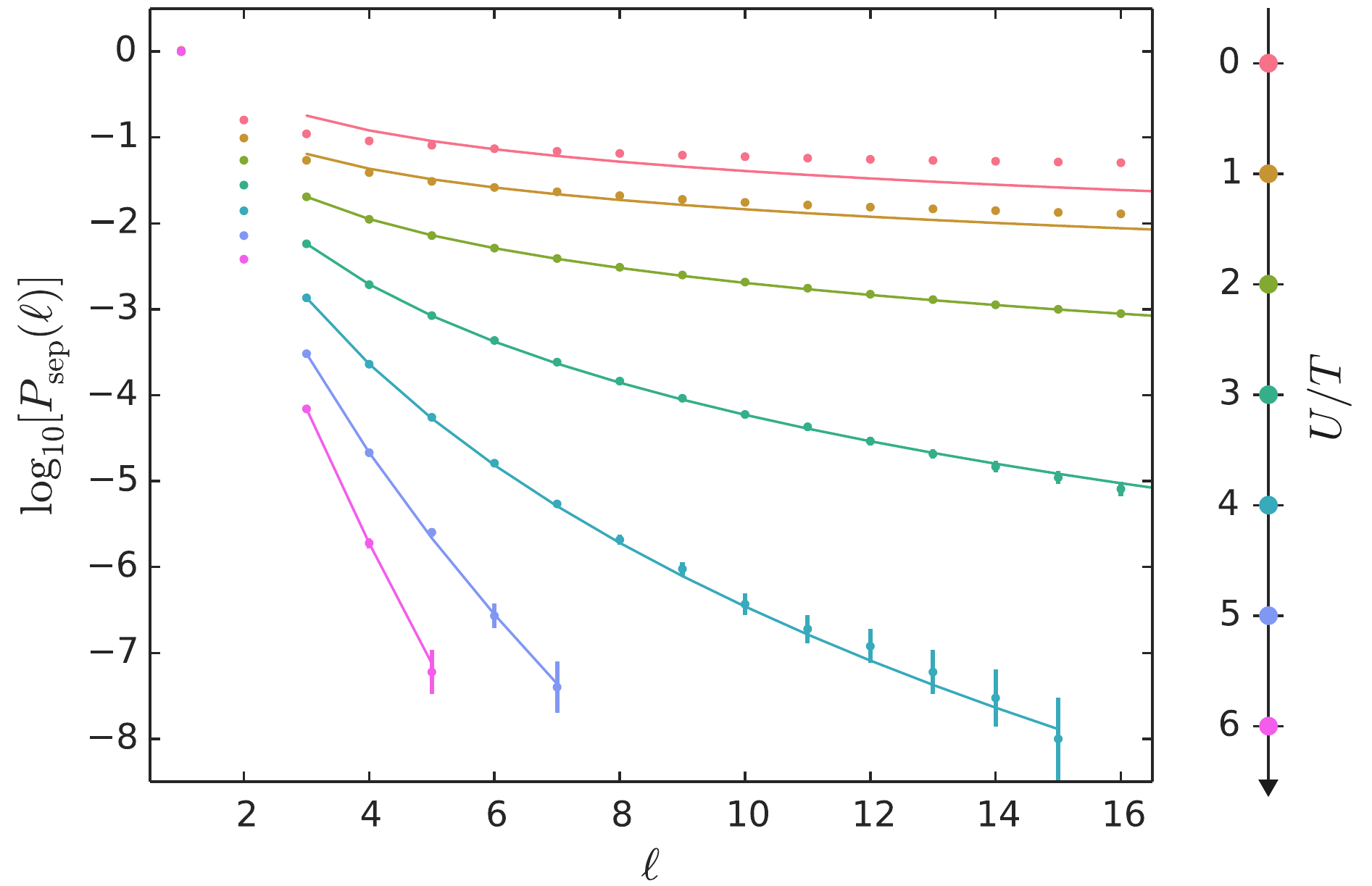}
    \caption{Probability that a single pair of anyons created on neighboring plaquettes separates by a maximum distance $\ell$ (in the direction relevant for logical errors) before recombining, shown for different potential depths $U/T$ and infinite (vanishing) potential-pump (-decay) rate. Data points and error bars are obtained by a kinetic Monte Carlo simulation of the pair dynamics (see Appendix~\ref{app:simulations}) with $10^8$ trajectories. Lines correspond to best fits based on our theoretical model [Eq.~\eqref{eq:1DResult}] with two fit parameters: $U$ and an overall prefactor. The corresponding values $U_\mathrm{fit}/T \approx 0.00$, $0.00$, $0.95$, $2.36$, $3.64$, $4.92$, $6.14$ (from top to bottom) agree very well for large $U/T$. Other relevant parameters are $\kappa = 1$, $T = 0.2$. The system size is set larger than the maximum observed pair separation. Data points for $\ell =1, 2$ are not used for the fits.}
    \label{fig:pair_separation_vs_U}
\end{figure}

Remarkably, our 1D model appears to capture the behavior of anyons beyond $\ell \approx \ell_\mathrm{1D}$ [Eq.~\eqref{eq:ell1D}]. Figure~\ref{fig:pair_separation_vs_U} shows a comparison of a kinetic Monte Carlo (KMC) simulation of the full dynamics (see Appendix~\ref{app:simulations}) with the estimate of Eq.~\eqref{eq:1DResult}. Even for moderate ratios $U/T \geq 2$, the data is very well described by the above 1D model (with renormalized values of $U$ for $U/T \leq 4$). Significant deviations are only observed for $U \lesssim T$ where we recover the behavior of free diffusion in 2D. In that case, the probability that an anyon pair separates by a distance $\ell$ decreases as $1/\log(\ell)$, as in the standard toric code~\cite{Brown2014_2}.

\subsection{Full dynamics of a single anyon pair}
 
So far, we have assumed that the effective potential induced by ancilla-plaquette interactions drops instantaneously to its minimum $-U$ on plaquettes visited by anyons, without recovery when anyons are absent. When considering the creation of multiple anyon pairs, this assumption would eventually lead to a situation where the entire system experiences a constant energy shift $-U$, in which case one would recover the dynamics of the standard toric code. To avoid this issue, it is thus important that the induced potential decays at a finite rate $\gamma_\mathrm{dec}$. In fact, in realistic physical implementations of the ancilla systems (see Appendix~\ref{app:ancillaSystems}), this potential not only decays at a finite rate, but also develops at a finite ``pump'' rate $\gamma_\mathrm{pump}$ when anyons are present. To recover the results of Sec.~\ref{sec:1DToyModel}, the pump and decay rates have to satisfy $\gamma_\mathrm{pump} \gg \gamma(0) \gg \gamma_\mathrm{dec}$. Below we discuss the effects of having such finite rates in more detail.

\subsubsection{Finite pump rate}
\label{sec:finitePump}

A finite pump rate makes it possible for anyons to move onto a different plaquette before a well-defined potential trench (with depth larger than $T$) can be established. Since this hinders the performance of our error-correction scheme, $\gamma_\mathrm{pump}$ should generally be as large as possible as compared to the free-diffusion rate $\gamma(0) = \kappa T$ [Eq.~\eqref{eq:transitionRate}]. To help suppress errors due to the finite pump rate, one could also consider ancilla systems that generate potential wells which extend by a few plaquettes around anyons (see Sec.~\ref{sec:conclusion}).

The effect of a finite pump rate is particularly relevant when an anyon ``jumps'' out of an existing potential trench. If $\gamma_\mathrm{pump} \gg \gamma(0)$, the anyon is very unlikely to diffuse further away from the trench before the latter is extended. In general, however, a finite time $\sim 1/\gamma_\mathrm{pump}$ will be required for it to sink in energy. During that time, the rate for the anyon to jump further away from the trench is given by $\gamma[U(t)]$ (where $U(t) = U[1-\exp(-\gamma_\mathrm{pump}t)]$ is the depth of the local potential at time $t$), which can be significantly larger than the rate $\gamma(U)$ obtained when the potential reaches its minimum. Since diffusion occurs on time scales longer than $1/\gamma_\mathrm{pump}$ in the regime of interest where $\gamma_\mathrm{pump}$ is the largest rate in the system, we estimate the probability to jump further away from the trench without extending it (i.e., to ``escape'' the trench) as $P_\mathrm{esc} = \int_0^{1/\gamma_\mathrm{pump}} \mathrm{d}t \, \gamma[U(t)]$. With $T/U \ll 1$, we obtain
\begin{equation}
    P_\mathrm{esc} \approx \frac{\pi^2}{6} \frac{T}{U} \frac{\gamma(0)}{\gamma_\mathrm{pump}}\,.
\end{equation}
Comparing $P_\mathrm{esc}$ with $\alpha_\uparrow = \gamma(U)/\gamma(0)$ then provides an estimate of the likelihood of ``trench escapes'' over ideal trench extensions.

\begin{figure}[t]
    \includegraphics[width=\columnwidth]{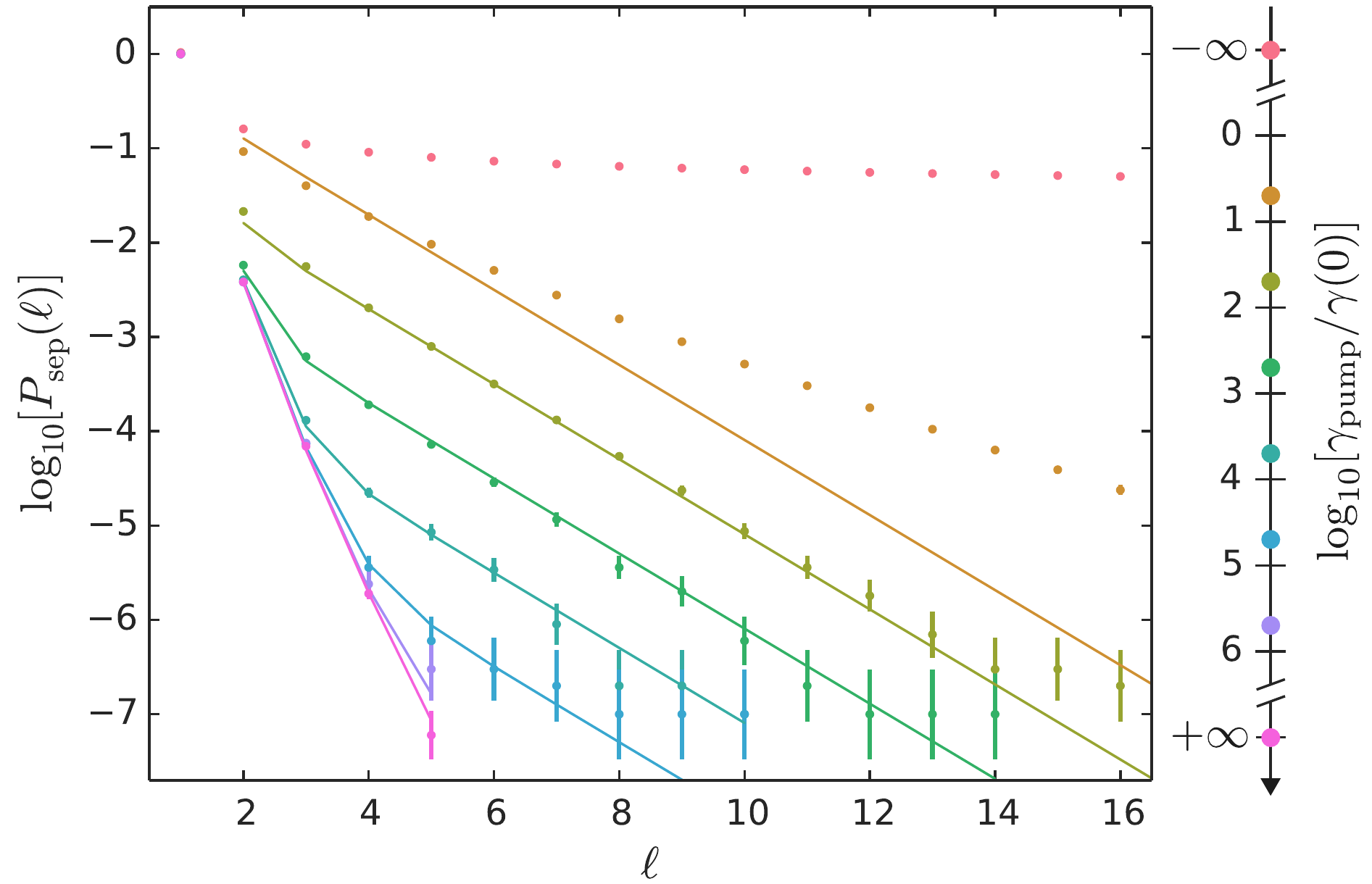}
    \caption{Pair-separation probability for different pump rates $\gamma_\mathrm{pump}$, with fixed potential depth $U/T = 6$ and vanishing potential decay rate. Data points and error bars are obtained through kinetic Monte Carlo simulation with $10^7$ trajectories. Lines correspond to fits given by $a P^{(0)}_\mathrm{sep}(\ell) + b P_\mathrm{esc} c^{\ell-2}$ with $(a,b,c) \approx (2.05, 2.25, 0.40)$, showing good agreement with the theoretical estimate of Eq.~\eqref{eq:1DResultFinitePump} (for large pump rates). Note that fitting is irrelevant for the uppermost curve (vanishing pump rate) which corresponds to the usual toric-code scaling. Other relevant parameters are $\kappa = 1$ and $T = 0.2$ as in Fig.~\ref{fig:pair_separation_vs_U} (the lowest curve is the same in both figures).}
    \label{fig:pair_separation_vs_pump} 
\end{figure}

The effect of a finite pump rate goes beyond increasing the probability of trench escapes. Once an anyon escapes and becomes separated from its original trench by a single plaquette $i$, a new potential trench most likely develops at the new anyon location. The probability that the anyon extends this new trench instead of going back to plaquette $i$ and joining the two trenches is then $P_2 \sim 1/2$. Intuitively, this stems from the fact that the potential barriers that the anyon must overcome to extend the new trench or come back to plaquette $i$ are similar ($\sim U$)~\footnote{Assuming that the anyon sinks to its potential minimum $-U$ after a (rare) escape event, the probability that it jumps even further from its original trench reads $\int_0^{1/\gamma_\mathrm{pump}} \mathrm{d}\tau (\gamma[U(\tau)]/P_\mathrm{esc})[\gamma(U)/(\gamma(U) + \gamma[U-U(\tau)])]$, which yields $P_2 = 1/4$ in the limit $U \gg T$}. To leading order,
the pair-separation probability $P_\mathrm{sep}(\ell)$ is thus increased by $P_\mathrm{esc}P_2^{\ell-2}$, namely,
\begin{equation}
    P_\mathrm{sep}(\ell) \approx P^{(0)}_\mathrm{sep}(\ell) + P_\mathrm{esc}P_2^{\ell-2} \, .
    \label{eq:1DResultFinitePump}
\end{equation}
Note that this behavior is still exponential, as confirmed by our KMC simulations (see Fig.~\ref{fig:pair_separation_vs_pump}). The exponential scaling persists provided that escape events are rare (i.e., $P_\mathrm{esc} < \alpha_\uparrow$, or $\log_{10}[\gamma_\mathrm{pump}/\gamma(0)] \gtrsim 1.3$ in Fig.~\ref{fig:pair_separation_vs_pump}).

\subsubsection{Finite decay rate --- trench splitting}
\label{sec:trenchSplitting}

\begin{figure}[t]
	\includegraphics[width=\columnwidth]{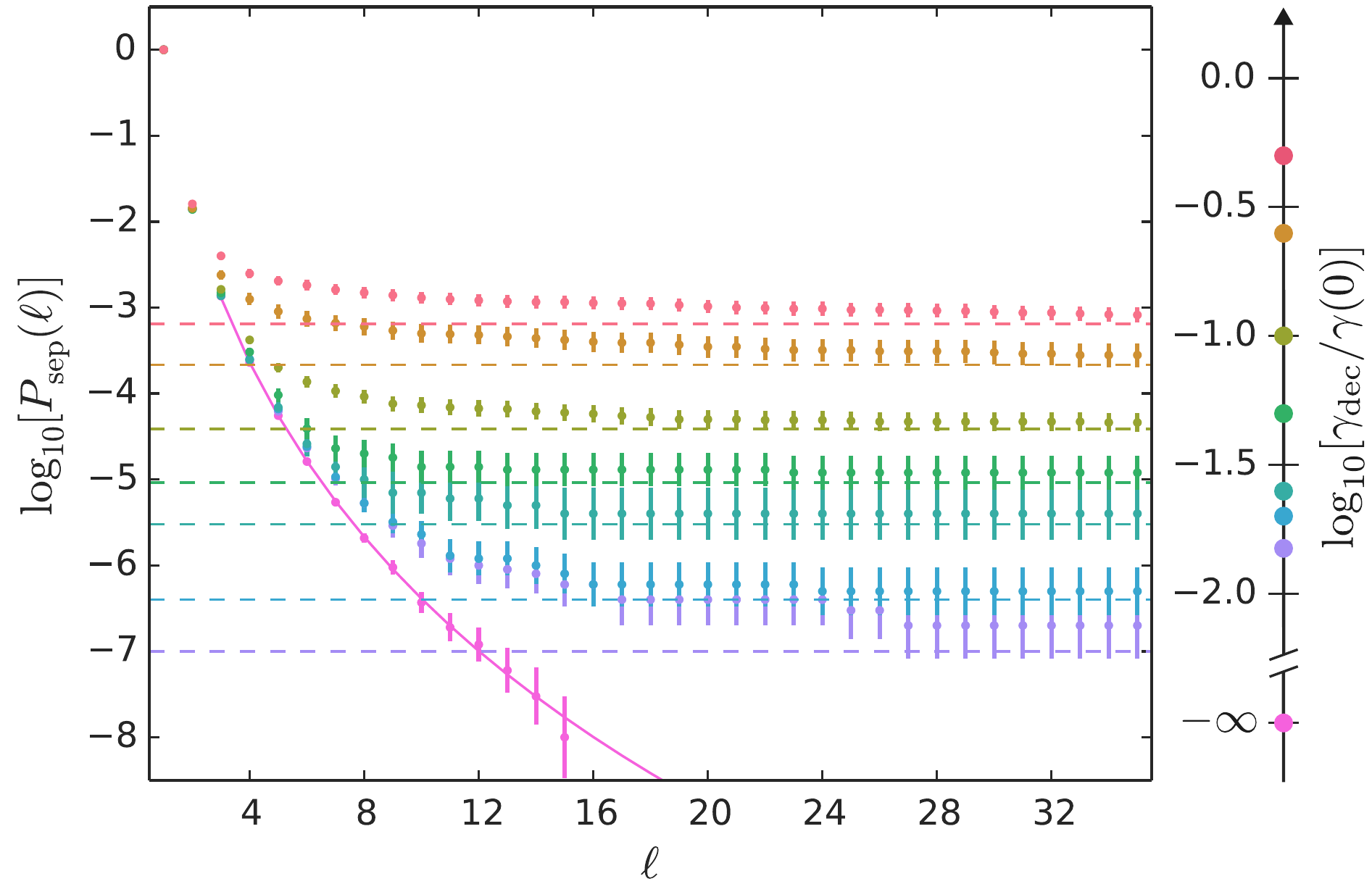}
	\caption{Pair-separation probability for different decay rates $\gamma_\mathrm{dec}$, with fixed potential depth $U/T = 4$ and infinite potential pump rate. Data points and error bars are obtained through kinetic Monte Carlo simulation with up to $10^8$ trajectories for the lowest set of points. The continuous curve shows a best fit of the form $aP^{(0)}_\mathrm{sep}(\ell)$ [Eq.~\eqref{eq:1DResult}] of the data points from $\ell = 4$ to $15$ corresponding to $\gamma_\mathrm{dec} = 0$, with $a$ and $U$ as fit parameters (yielding $U_\mathrm{fit}/T \approx 3.58$, in good agreement with the actual value). The dashed lines indicate the value at which $P_\mathrm{sep}(\ell)$ saturates for each non-zero value of $\gamma_\mathrm{dec}$. The crossing between each of these lines and the curve corresponding to $\gamma_\mathrm{dec} = 0$ provides an estimate of the length at which saturation occurs: $\ell_\mathrm{max, fit} \approx 3.4$, $4.0$, $5.3$, $6.5$, $7.6$, $10.0$, $12.0$ (from top to bottom); in remarkable agreement, for large $\ell_\mathrm{max}$, with the corresponding theoretical estimates $\ell_\mathrm{max} = 2.0$, $2.8$, $4.5$, $6.3$, $8.9$, $10.0$, $11.5$ [see Eq.~\eqref{eq:ellMax}]. Other relevant parameters are $\kappa = 1$ and $T = 0.2$ as in Figs.~\ref{fig:pair_separation_vs_U} and~\ref{fig:pair_separation_vs_pump}. Dashed lines are obtained using points for $\ell = 50$ to $120$ (not shown).}
	\label{fig:pair_separation_vs_decay}
\end{figure}

We have argued that a finite potential decay rate $\gamma_\mathrm{dec}$ is required to ensure that potential trenches do not cover the whole system eventually. For a single pair in a single trench, the decay also makes it possible for the trench to split into two separate pieces once the anyon separation becomes large enough. Indeed, a trench can only be stable if all of its plaquettes can be visited within a time $t_\mathrm{d} \approx (T/U) \gamma_\mathrm{dec}^{-1}$ which corresponds to the time required for local potentials to decay by at most $T$ from their initial value $\approx -U$. Since the mean square deviation for the (essentially free) diffusion within the trench during that time is given by $r^2 = 2 \eta Dt_\mathrm{d}$ (where $D \equiv \gamma(0)$ is the diffusion constant and $1 \geq \eta \geq 2$ is the effective dimensionality of the trench), we can estimate the maximum size $\ell_\mathrm{max} = 2r$ that can be reached by a single trench as
\begin{equation}
    \ell_\mathrm{max} \approx 2\sqrt{2 \eta} \sqrt{\frac{T}{U} \frac{\gamma(0)}{\gamma_\mathrm{dec}}}\,.
    \label{eq:ellMax}
\end{equation}
For system sizes larger than $2\ell_\mathrm{max}$, anyons from the same pair which separate by $\ell \sim \ell_\mathrm{max}$ therefore likely end up trapped in distinct potential trenches. Since at that point the system has no means to remember that the anyons originated from the same pair, each anyon subsequently performs a free 2D diffusion as in the standard toric code (with a reduced diffusion constant since thermal activation is required to extend individual trenches). Consequently, the exponential decrease of the separation probability [Eq.~\eqref{eq:1DResult}; or~\eqref{eq:1DResultFinitePump} when $\gamma_\mathrm{pump}$ is finite] saturates at about $P_\mathrm{sep}(\ell_\mathrm{max})$ for large system sizes. Although this introduces an upper limit to the performance of our error-correction scheme, we emphasize that the exponential decay of $P_\mathrm{sep}(\ell)$ up to $\ell \sim \ell_\mathrm{max}$ can lead to pair-separation probabilities (and therefore error probabilities) orders of magnitude smaller than in the usual toric code.

We have performed KMC simulations to verify the validity of our estimate for $\ell_\mathrm{max}$ [Eq.~\eqref{eq:ellMax}]. As illustrated in Fig.~\ref{fig:pair_separation_vs_decay}, our results demonstrate a good agreement with our theoretical model. As expected, the initial exponential decay of $P_\mathrm{sep}(\ell)$ crosses over to a plateau for $\ell \gtrsim \ell_\mathrm{max}$.

\section{Memory figures of merit}
\label{sec:memoryFiguresOfMerit}

Building on our understanding of the dynamics of a single anyon pair, we now investigate the average coherence time of the quantum memory in the more realistic scenario where multiple anyon pairs can be created. As in the standard toric code (see, e.g., Ref.~\cite{Brown2014_2}), the coherence time $\tau_\mathrm{coh}$ consists of two contributions $\tau_\mathrm{cre} + \tau_\mathrm{sep}$, where $\tau_\mathrm{cre}$ is the average waiting time until the creation of an error-causing pair of anyons, and $\tau_\mathrm{sep}$ is the average time required for the anyons of such a pair to actually separate and cause an error. In contrast to the situation obtained in the standard toric code where it is typically enough to create a single pair of anyons to cause an error, here $\tau_\mathrm{cre}$ accounts for the fact that many attempts are required to create an error-causing pair:
\begin{equation}
    \tau_\mathrm{cre}^{-1} \approx 2 L^2 \gamma(2J) P_\mathrm{sep}(L/2),
    \label{eq:tau_cre}
\end{equation}
where $2 L^2\gamma(2J)$ is the anyon-pair-creation rate (assuming that potential trenches decay much faster than the typical time between creation events), and $P_\mathrm{sep}(L/2)$ is the probability that anyons of a pair separate by half the system size, thus causing a logical error.

The separation time $\tau_\mathrm{sep}$ can estimated as the time required for the anyons of a pair to diffuse away from each other by a distance $L/2$ via thermally-activated steps with activation energy $\sim U$. An upper bound can be obtained by assuming that each diffusion step is thermally activated with rate $\sim \gamma(U)$, yielding $\tau_\mathrm{sep} \lesssim L^2/\gamma(U)$~\footnote{The rate at which a potential trench of size $\ell$ is extended is $\sim \gamma(U)/\ell$, where $1/\ell$ is the approximate probability that an anyon is located at the trench-potential wall. The total time required to extend a trench from initial size $2$ to $L/2$ is thus $\tau_\mathrm{sep} \sim \sum_{\ell = 2}^{L/2} \gamma(U)^{-1} \ell\sim \gamma(U)^{-1} L^2$, yielding $\tau_\mathrm{sep} \sim \gamma(U)/L^2$ (or $\sim \gamma(U)/L^3$ for 2D trenches).}. As compared to the exponential or fast power-law increase of $\tau_\mathrm{cre}$ with $L$ [Eqs.~\eqref{eq:tau_cre} and~\eqref{eq:1DResultFinitePump}], the contribution of $\tau_\mathrm{sep}$ to the coherence time can thus safely be neglected (in the low-temperature regime $T \ll U$). Therefore, $\tau_\mathrm{coh} \approx \tau_\mathrm{cre}$. In the following, we identify the leading error sources arising from the possibility of having multiple anyon pairs and provide a qualitative estimate of the maximum achievable coherence time.

\subsection{Effects of multiple anyon-pair-creation events}

We have shown in Sec.~\ref{sec:singleError} that the main limitation to our scheme arises, for a single anyon pair, from the finite decay rate of potential trenches. This decay introduces a length scale $\ell_\mathrm{max}$ [Eq.~\eqref{eq:ellMax}] beyond which the anyon-pair-separation probability $P_\mathrm{sep}(\ell)$ is expected to saturate. In the general situation where additional anyon pairs can be created by absorbing energy from the bath, this non-zero decay rate becomes required. Indeed, to ensure that our error-correction mechanism remains active, potential trenches left by anyons that annihilate must be ``erased'' fast enough before new anyon pairs are created. This emphasizes the importance of dissipation in our scheme.

The creation of a new anyon pair can either occur inside an existing potential trench or lead to the creation of a new potential trench~\footnote{Anyon pairs can also be created partly inside an existing trench with rate $\sim \ell \gamma(2J-U)$. Such events can be neglected as compared to anyon-pair creations inside a trench, which occur at a faster rate [in the low-temperature regime of interest where $T \ll U$, such that $\gamma(2J-U) \ll \gamma(2J-2U)$].}. Anyon-pair creation in an existing trench occurs with a rate $\ell^\eta \gamma(2J-2U)$, where $\ell^\eta$ is the area of the existing trench [with length scale $\ell$ and effective dimensionality $1 \leq \eta \leq 2$; see Eq.~\eqref{eq:ell1D}]. Two situations can be distinguished depending on whether the trench is on average occupied or not when a new anyon pair is created. We refer to them as ``trench saturation'' and ``refilling'', respectively. As we demonstrate below, trench refilling imposes a lower bound on the potential-decay rate $\gamma_\mathrm{dec}$. More importantly, trench saturation leads to an upper bound $L_\mathrm{max}^\mathrm{sat}$ beyond which increasing the system size is not expected to further enhance the memory-coherence time. Although this maximum does not depend on $\gamma_\mathrm{dec}$, keeping our scheme effective up to a system size $L_\mathrm{max}^\mathrm{sat}$ requires a suitable choice of $\gamma_\mathrm{dec}$, due to additional trench-refilling and trench-splitting effects.

Anyon-pair creation away from any existing trench occurs with a rate $\sim L^2 \gamma(2J)$ (where $L$ denotes the system size). It generates new independent trenches, which can be harmful if the latter join and percolate to create larger trenches of size $\sim L/2$. As we demonstrate in Appendix~\ref{app:trenchPercolation}, however, trench percolation can be neglected when $\gamma_\mathrm{dec} \gg \gamma(2J)$, which is automatically satisfied under the requirements imposed by trench refilling.

\subsubsection{Trench saturation}
\label{sec:trenchSaturation}

The creation of new anyon pairs inside an existing potential trench crucially modifies its average anyon density $\rho_\mathrm{t}$. If anyon creation dominates over anyon recombination, more than one anyon is found in the trench on average. In that case, the trench effectively does not decay, and the probability that it becomes larger ultimately saturates instead of decreasing exponentially with trench size $\ell$ with no bound. As for trench-splitting effects discussed in Sec.~\ref{sec:trenchSplitting}, this introduces a maximum system size beyond which we do not expect our scheme to further enhance the memory-coherence time. To estimate this upper bound $L_\mathrm{max}^\mathrm{sat}$, we determine the average density of anyons in a trench in a mean-field self-consistent way: For a finite density $\rho_\mathrm{t}$, the average distance between anyons is $\delta_\mathrm{t} \sim \rho_\mathrm{t}^{-1/\eta}$. New anyons separated by such a distance are created with a rate $\gamma \sim \ell^\eta \gamma(2J-2U) P_\mathrm{s}(\delta_\mathrm{t})$, where $P_\mathrm{s}(\delta_\mathrm{t}) \sim 1/\delta_\mathrm{t}^{\eta'}$ is the probability that anyons created as a local pair separate by a distance $\delta_\mathrm{t}$ instead of recombining, with $0 < \eta' \leq 1$ depending on the effective dimensionality of potential trenches~\footnote{The probability that diffusing anyons separate by a distance $\ell$ without recombining is $\sim 1/\ell$ in 1D [see above Eq.~\eqref{eq:1DResult}], and $\sim 1/\log(\ell)$ in 2D (see, e.g., Ref.~\cite{Brown2014_2}). We neglect non-essential corrections coming from the bias towards annihilation when two anyons meet.}. Since anyons diffuse and most likely annihilate when they meet each other, their average lifetime is $\tau \sim \delta_\mathrm{t}^2/\gamma(0)$. Therefore, the average anyon density in the trench should satisfy $\rho_\mathrm{t} = \gamma \tau/\ell^\eta$, which yields
\begin{equation}
    \rho_\mathrm{t} \sim \left[\frac{\gamma(2J-2U)}{\gamma(0)}\right]^{\eta/(2+\eta-\eta')}.
    \label{eq:rhot}
\end{equation}
As expected, $\rho_\mathrm{t}$ is independent of the trench size $\ell$, and increases with larger trench depth $U < J$.

As discussed above, trench saturation occurs when, on average, more than one anyon occupies the trench, i.e., $\rho_\mathrm{t} \gtrsim 1/\ell^\eta$. Small values $\rho_\mathrm{t} \ll 1/\ell^\eta$ indicate that saturation is irrelevant (i.e., that the trench is most likely empty when new anyon pairs are created, or that it has already fully decayed). Using Eq.~\eqref{eq:rhot}, we find that saturation occurs at system sizes larger than
\begin{equation}
    L_\mathrm{max}^\mathrm{sat} \approx 2 \delta_\mathrm{t} \sim 2 \left[\frac{\gamma(0)}{\gamma(2J-2U)}\right]^{1/(2+\eta-\eta')}.
    \label{eq:LMaxTrenchSaturation}
\end{equation}
%

\subsubsection{Trench refilling}
\label{sec:trenchRefilling}

Trench saturation is suppressed when the system size satisfies $L \lesssim L_\mathrm{max}^\mathrm{sat}$. In that case, potential trenches can end up empty and gradually disappear, as desired. Below, we quantify the minimum potential-decay rate $\gamma_\mathrm{dec}$ required for this process to be effective.

Anyons that are created in an existing empty trench before the latter has sufficiently decayed can diffuse quasi-freely in the trench and ``reactivate'' the latter. This effect, which we call ``trench refilling'', is only relevant when new anyon pairs are created on a faster time scale than the time $t_\mathrm{d} \sim (T/U)\gamma_\mathrm{dec}^{-1}$ required for the empty trench to decay by $\Delta U \gtrsim T$. In that case, anyons can diffuse quasi-freely in the old trench for a time $\sim t_\mathrm{d}$, which allows them to reach distances of the order of $d \sim (2\eta \gamma(0) t_\mathrm{d})^{1/2} \approx \ell_\mathrm{max}/2$, i.e., to span the entire existing trench [see Eq.~\eqref{eq:ellMax}]. Although the probability to form an original trench of size $\ell$ may have been exponentially small (in $\ell$), the probability that the new anyons separate by a distance $\ell$ (thereby ``reactivating'' the trench) is now $P_\mathrm{sep}^\mathrm{new}(\ell) \sim 1/\ell^{\eta'}$. The average number of such reactivations is $n_\mathrm{r}(\ell) \sim \ell^\eta \gamma(2J-2U)P_\mathrm{sep}^\mathrm{new}(\ell)t_d$, where $\ell^\eta \gamma(2J-2U)$ is the rate for anyon-pair creation inside the trench. To suppress the effects of trench refilling, we require that $n_\mathrm{r}(\ell) \ll 1$ for any trench size $\ell \lesssim L/2$, where $L$ is the system size. As anticipated above, this provides a minimum for the trench-decay rate:
\begin{equation}
    \gamma_\mathrm{dec} \gg \frac{T}{U} \left(\frac{L}{2}\right)^{\eta-\eta'} \gamma(2J-2U).
    \label{eq:gammaDecLowerBound1}
\end{equation}
We remark that this lower bound is independent of $L$ for 1D trenches ($\eta = \eta' = 1$), whereas it essentially scales as $L^2$ for 2D trenches ($\eta = 2$ and $\eta' \rightarrow 0$)~\footnote{As mentioned above Eq.~\eqref{eq:rhot}, $(L/2)^{\eta-\eta'}$ should be replaced by $(L/2)^2/\log(L/2)$ for 2D trenches.}. Note that we have implicitly assumed that $\gamma_\mathrm{pump}/\gamma(0) \gg 1$ in deriving Eq.~\eqref{eq:gammaDecLowerBound1}. The potential-pump rate determines the rate at which new anyons can develop a new trench in the old one. If it is reduced away from the above ideal limit, the self-trapping of new anyon pairs becomes less efficient, which has two effects: First, it increases the time $t_\mathrm{d}$ over which the new anyons can diffuse quasi-freely in the existing trench. Second, it increases the probability $P_\mathrm{sep}^\mathrm{new}(\ell)$ that they separate by a distance $\ell$. Both of these effects increase the number of reactivations of empty trenches, as we demonstrate numerically in Sec.~\ref{sec:simulationsCoherenceTime}.

\subsection{Estimate of the maximum coherence time}
\label{sec:maxCoherenceTime}

We have demonstrated that trench-splitting and refilling effects lead to upper and lower bounds on the trench-decay rate, respectively. Summarizing our results from Eqs.~\eqref{eq:ellMax} and~\eqref{eq:gammaDecLowerBound1}, we obtain the following requirements:
\begin{equation}
    \gamma(2J-2U) \frac{T}{U} \left(\frac{L}{2}\right)^{\eta-\eta'} \!\!\!\!\!\! \ll \gamma_\mathrm{dec} \ll 2 \eta \gamma(0) \frac{T}{U} \left(\frac{L}{4}\right)^{-2} \!\!\!\!\!,
    \label{eq:boundsOnGammaDec}
\end{equation}
which apply in the low-temperature regime $T \ll U,\, J-U$ (with $J > U$), under the assumption that $\gamma_\mathrm{pump} \gg \gamma(0)$ [we recall that $\gamma(0) \sim T$ and $\gamma(2J-2U) \sim 2(J-U) \mathrm{e}^{-2(J-U)/T}$; see Eq.~\eqref{eq:transitionRate}]. This key result illustrates the power and limitations of our scheme: First, it shows that $T > 0$ is crucially required ($\gamma(0)$ decreases with temperature, which makes the right inequality harder to satisfy with decreasing $T$), which emphasizes the importance of diffusion in our scheme. Second, Eq.~\eqref{eq:boundsOnGammaDec} highlights the fact that increasing the trench-energy scale $U$ is not only beneficial: Although $U \gg T$ is required, increasing $U$ enhances trench-refilling and splitting effects, making both sides of Eq.~\eqref{eq:boundsOnGammaDec} harder to satisfy. Trench refilling can be efficiently suppressed, however, by ensuring that $J-U \gg T$ (i.e., by increasing the energy gap $J$ of the toric-code system). Finally, Eq.~\eqref{eq:boundsOnGammaDec} confirms the existence of a maximum system size $L_\mathrm{max}$ beyond which the memory-coherence time should not improve further. Here, this maximum can be estimated by equating both sides of Eq.~\eqref{eq:boundsOnGammaDec}, yielding
\begin{eqnarray}
    L_\mathrm{max} & \sim & 2 \left[ 8\eta \frac{\gamma(0)}{\gamma(2J-2U)} \right]^{1/(2+\eta-\eta')} \nonumber \\
    & \approx & 2 \left[ 4\eta \frac{T}{J-U} \mathrm{e}^{2(J-U)/T} \right]^{1/(2+\eta-\eta')}.
    \label{eq:LMaxFromBounds}
\end{eqnarray}
Remembering the results of Sec.~\ref{sec:trenchSaturation}, we notice that $L_\mathrm{max} = 2(8\eta)^{1/(2+\eta-\eta')} \delta_t \sim L_\mathrm{max}^\mathrm{sat}$. Therefore, the maximum $L_\mathrm{max}$ resulting from the requirements due to trench-refilling and splitting effects essentially coincides with the upper bound $L_\mathrm{max}^\mathrm{sat}$ due to trench saturation [Eq.~\eqref{eq:LMaxTrenchSaturation}]. The most important feature of $L_\mathrm{max}$ is that it increases exponentially with $(J-U)/T \gg 1$. To keep our scheme effective up to this maximum system size, the following optimal trench-decay rate is expected to be required:
\begin{equation}
    \gamma_\mathrm{dec}^\mathrm{opt} \sim 2\eta \gamma(0) \frac{T}{U} \left[ \eta \frac{T}{J-U} \mathrm{e}^{2(J-U)/T} \right]^{-2/(2+\eta-\eta')},
    \label{eq:optDecayRate}
\end{equation}
which also scales exponentially with $(J-U)/T$. Therefore, our scheme can be effective up to very large system sizes provided that the potential-decay rate can be made sufficiently small. As long as the requirements given by Eq.~\eqref{eq:boundsOnGammaDec} are satisfied, the memory-coherence time should increase exponentially with system size, which is one of the main results of this work. To estimate the maximum coherence time that can in principle be achieved, one can introduce $L = L_\mathrm{max}$ in Eq.~\eqref{eq:tau_cre}, thereby obtaining
\begin{eqnarray}
    \tau_\mathrm{coh}^\mathrm{max} & \sim & [2\gamma(2J)]^{-1} L_\mathrm{max}^{-2} P^{-1}_\mathrm{sep}(L_\mathrm{max}/2) \nonumber \\
    & \sim & \left(\mathrm{e}^{c_1 \beta J}\right)^{\mathrm{e}^{c_2 \beta U}},
    \label{eq:maxCoherenceTime}
\end{eqnarray}
where $c_1$ and $c_2$ are positive numbers of order $1$. In the specific scenario where $J > 3U$, such that $\alpha_\uparrow L_\mathrm{max} > 1$, one finds $c_1 = (J-3U)/J$ and $c_2 = 1$ [using Eqs.~\eqref{eq:1DResult} and~\eqref{eq:LMaxFromBounds}]. Therefore, due to the exponential scaling of $L_\mathrm{max}$ with inverse temperature $\beta$, the memory-coherence time can increase according to a double-exponential scaling, in stark contrast to the Arrhenius law $\tau_\mathrm{coh} \sim \mathrm{e}^{2\beta J}$ typically observed in quantum memories protected by a gap $2J$. Eq.~\eqref{eq:maxCoherenceTime} tells us that the maximum coherence time increases with $U/T$~\footnote{In the regime of interest where $(J-U)/T \gg 1$, $\tau_\mathrm{coh}^\mathrm{max}$ increases with $U/T$ despite the exponential factor $\mathrm{e}^{\beta(J-U)/2}$ in the double exponential.} despite the fact that the system size $L_\mathrm{max}$ up to which our scheme is effective \emph{decreases}. We demonstrate this behavior numerically in Sec.~\ref{sec:simulationsCoherenceTime} (Fig.~\ref{fig:coherence_time_vs_U}).

If $\gamma_\mathrm{dec}$ cannot be made small enough to reach the optimal value of Eq.~\eqref{eq:optDecayRate}, the left inequality of Eq.~\eqref{eq:boundsOnGammaDec} (governing trench-refilling effects) becomes irrelevant, and trench-splitting effects (governed by the right inequality) become the main limitation. For a fixed $\gamma_\mathrm{dec}$, the maximum system size thus becomes
\begin{eqnarray}
    L_\mathrm{max}' & \sim & \ell_\mathrm{max} \approx 2\sqrt{2 \eta} \sqrt{\frac{T}{U} \frac{\gamma(0)}{\gamma_\mathrm{dec}}},
    \label{eq:LMaxPrime}
\end{eqnarray}
which coincides with the maximum size $\ell_\mathrm{max}$ that can be reached by a potential trench with a single anyon pair [see Eq.~\eqref{eq:ellMax}]. Clearly, $\gamma(0)/\gamma_\mathrm{dec} \gg U/T$ must be ensured in that case. This can be achieved, e.g., by increasing the bath-coupling constant $\kappa = \gamma(0)/T$ [see Eq.~\eqref{eq:transitionRate}]. Eq.~\eqref{eq:LMaxPrime} leads to the more conservative estimate
\begin{eqnarray}
    \tau_\mathrm{coh} & \sim & [2\gamma(2J)]^{-1} \ell_\mathrm{max}^{-2} P^{-1}_\mathrm{sep}(\ell_\mathrm{max}/2) \nonumber \\
    & \sim & \frac{\mathrm{e}^{2\beta J}}{\ell_\mathrm{max}^2} \left(\frac{1}{\beta U} \frac{\mathrm{e}^{\beta U}}{\ell_\mathrm{max}}\right)^{\ell_\mathrm{max}},
    \label{eq:maxCoherenceTime2}
\end{eqnarray}
where we have assumed that $\ell_\mathrm{max} \ll \mathrm{e}^{\beta U}$ (note that when $\ell_\mathrm{max}$ becomes of the order of $\mathrm{e}^{\beta U}$, which requires an exponentially small $\gamma_\mathrm{dec}$, we recover the double-exponential scaling discussed above). We remark that the exponent $\beta U \ell_\mathrm{max}$ in Eq.~\eqref{eq:maxCoherenceTime2} is independent of temperature. In summary, in the above scenario which does not take full advantage of our scheme (leading to an Arrhenius law), the coherence time is still significantly improved --- by an exponentially large temperature-independent prefactor.

\subsection{Simulations of the memory-coherence time}
\label{sec:simulationsCoherenceTime}

In this section, we present the results of our numerical investigation of the memory-coherence time under the full dynamics of the system, including all effects discussed in the previous sections. We demonstrate that our theoretical arguments and estimates are consistent with these results, and provide additional insight into the efficiency of our scheme. We remark that all ideal parameter regimes in which our error-correction mechanism is most efficient cannot be explored with a reasonable amount of computing resources. Therefore, in the following, we deliberately choose non-optimal parameters where effects such as trench splitting limit the exponential improvement of the coherence time.

\begin{figure}[t]
	\includegraphics[width=\columnwidth]{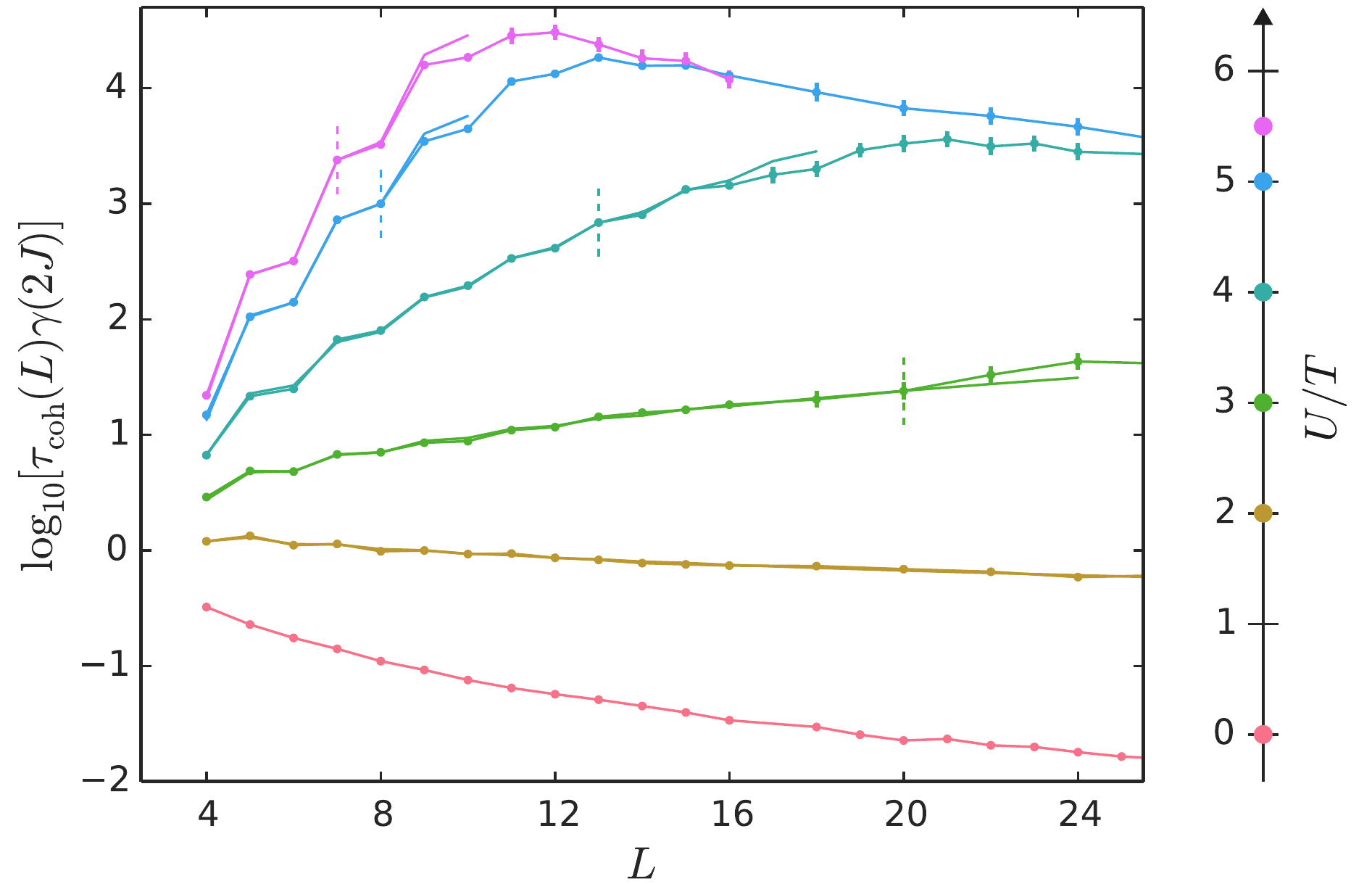}
	\caption{Coherence time as a function of system size $L$, for different maximum potential depths $U/T$. The trench-decay rate is set to $10^{-3}\gamma(0)$, except for the upper two curves where it is chosen 10 times larger to make it closer to the expected ``optimal'' value [Eq.~\eqref{eq:optDecayRate}]. The potential-pump rate is effectively infinite, and data points and error bars are obtained by a kinetic Monte Carlo simulation of the full system dynamics with $10^3$ trajectories. The upper five curves are fitted based on our theoretical model [Eqs.~\eqref{eq:1DResult} and~\eqref{eq:tau_cre}] using the largest number of points that can provide a reasonable fit (according to a chi-square goodness-of-fit test; using $U$ and a prefactor as fitting parameters). Dashed vertical bars delimit the corresponding fitting range (starting at $L = 4$), and fitting curves are extended slightly beyond to illustrate where deviations occur. Despite the trench-decay rate and the mostly 2D nature of potential trenches (see the text), we find $U_{\mathrm{fit}} = 5.33$, $4.82$, $3.51$, $2.20$, $0.67$, $0.00$, in good agreement with our model (for large $U/T$). The obtained values of $U/T$ are consistently smaller than the actual ones due to the 2D nature of potential trenches and to the finite decay rate which effectively reduces the potential depth. Although the decay rate is not supposed to be optimal (for all values $U/T$), the coherence time nevertheless saturates at system sizes $L \approx L_\mathrm{max}$. The crossovers observed for the upper three curves are in remarkable agreement with $L_\mathrm{max} \approx 11.0$, $13.7$, $21.6$ obtained from Eq.~\eqref{eq:LMaxTrenchSaturation}.}
	\label{fig:coherence_time_vs_U}
\end{figure}

\begin{figure}[t]
	\includegraphics[width=\columnwidth]{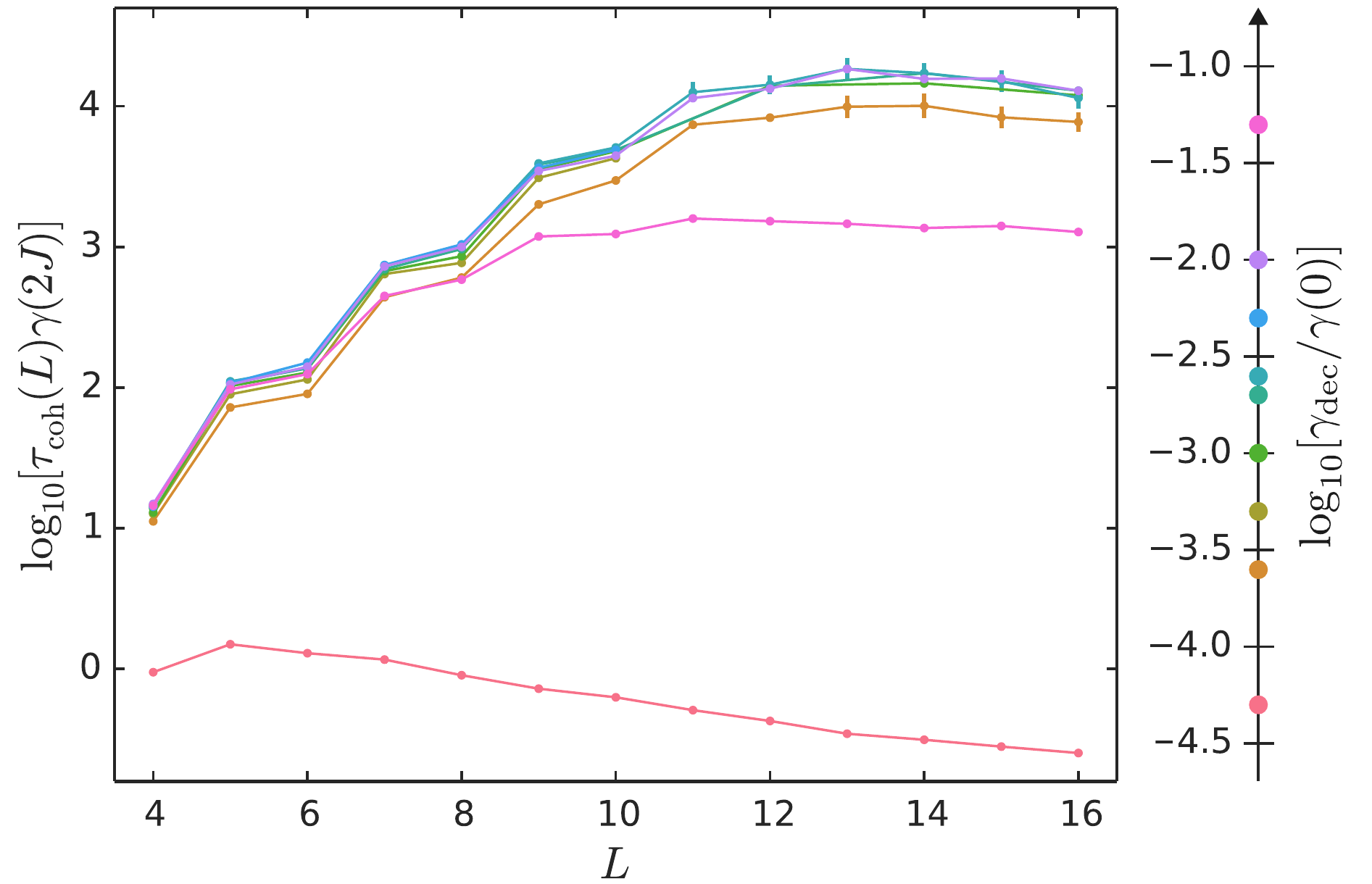}
	\caption{Coherence time as a function of system size, for different potential-decay rates $\gamma_\mathrm{dec}$ and fixed $U/T = 5$. The potential-pump rate is effectively infinite, and data points and error bars are obtained through a kinetic Monte Carlo simulation of the full system dynamics with $10^3$ trajectories. Although all curves (except the bottom one) can be reproduced using Eqs.~\eqref{eq:tau_cre} and~\eqref{eq:1DResultFinitePump} as in Figs.~\ref{fig:coherence_time_vs_U} and~\ref{fig:coherence_time_vs_pump}, no fitting curve is shown here, for clarity. Remarkably, variations of the potential-decay rate over several orders of magnitude essentially do not affect the coherence-time scaling. All curves in this parameter range saturate at $L_\mathrm{max} \approx 12$, in reasonable agreement with $L _\mathrm{max} \approx 13.7$ obtained from our theoretical model [Eq.~\eqref{eq:LMaxTrenchSaturation}]. The fact that saturation at this maximum is reached for decay rates away from the estimated ``optimal'' value $\log_{10}[\gamma_\mathrm{dec}^\mathrm{opt}/\gamma(0)] \approx -3.2$ [Eq.~\eqref{eq:optDecayRate}] indicates that trench-refilling and splitting effects impose softer constraints on our scheme than trench saturation, as discussed in the text.}
	\label{fig:coherence_time_vs_decay}
\end{figure}

We first illustrate, in Fig.~\ref{fig:coherence_time_vs_U}, the significant change of behavior induced by potential trenches. When the maximum depth of trenches is negligible as compared to temperature ($U/T \ll 1$), the memory-coherence time decreases as $\tau_\mathrm{coh} \sim \log(L/2)/L^2$ with system size $L$, which corresponds to the usual behavior of the toric code~\cite{Brown2014_2}. When the potential-trench energy scale $U$ reaches values $U \gtrsim T$, however, a clear exponential increase develops, in agreement with our theoretical estimates. In Fig.~\ref{fig:coherence_time_vs_U}, this exponential scaling saturates and crosses over to a $\sim \log(L/2)/L^2$ behavior at relatively small system sizes, which stems from the non-optimal choice of parameters mentioned above. We recall that the expected maximum $L_\mathrm{max}$ corresponds to the threshold at which trench-saturation, refilling and splitting effects simultaneously come into play [Eqs.~\eqref{eq:LMaxTrenchSaturation} and~\eqref{eq:LMaxFromBounds}], and that we expect to require an optimal value of the trench-decay rate $\gamma_\mathrm{dec}$ to reach it [Eq.~\eqref{eq:optDecayRate}]. Here, the observed crossovers are consistent with our theoretical estimates for $L_\mathrm{max}$ [Eqs.~\eqref{eq:LMaxTrenchSaturation} and~\eqref{eq:LMaxFromBounds}] despite the fact that the trench-decay rate $\gamma_\mathrm{dec}$ does not precisely corresponds to the estimated optimal value. This illustrates an important property of our scheme, namely, its robustness with respect to tuning $\gamma_\mathrm{dec}$. We remark that potential trenches are mostly two-dimensional in the parameter regimes explored in this section, such that $\eta - \eta' \approx 2$ in Eq.~\eqref{eq:LMaxTrenchSaturation}. The requirement for purely 1D trenches [Eq.~\eqref{eq:ell1D}] is not satisfied, even for the largest values of $U/T$ used in our simulations.

\begin{figure}[t]
	\includegraphics[width=\columnwidth]{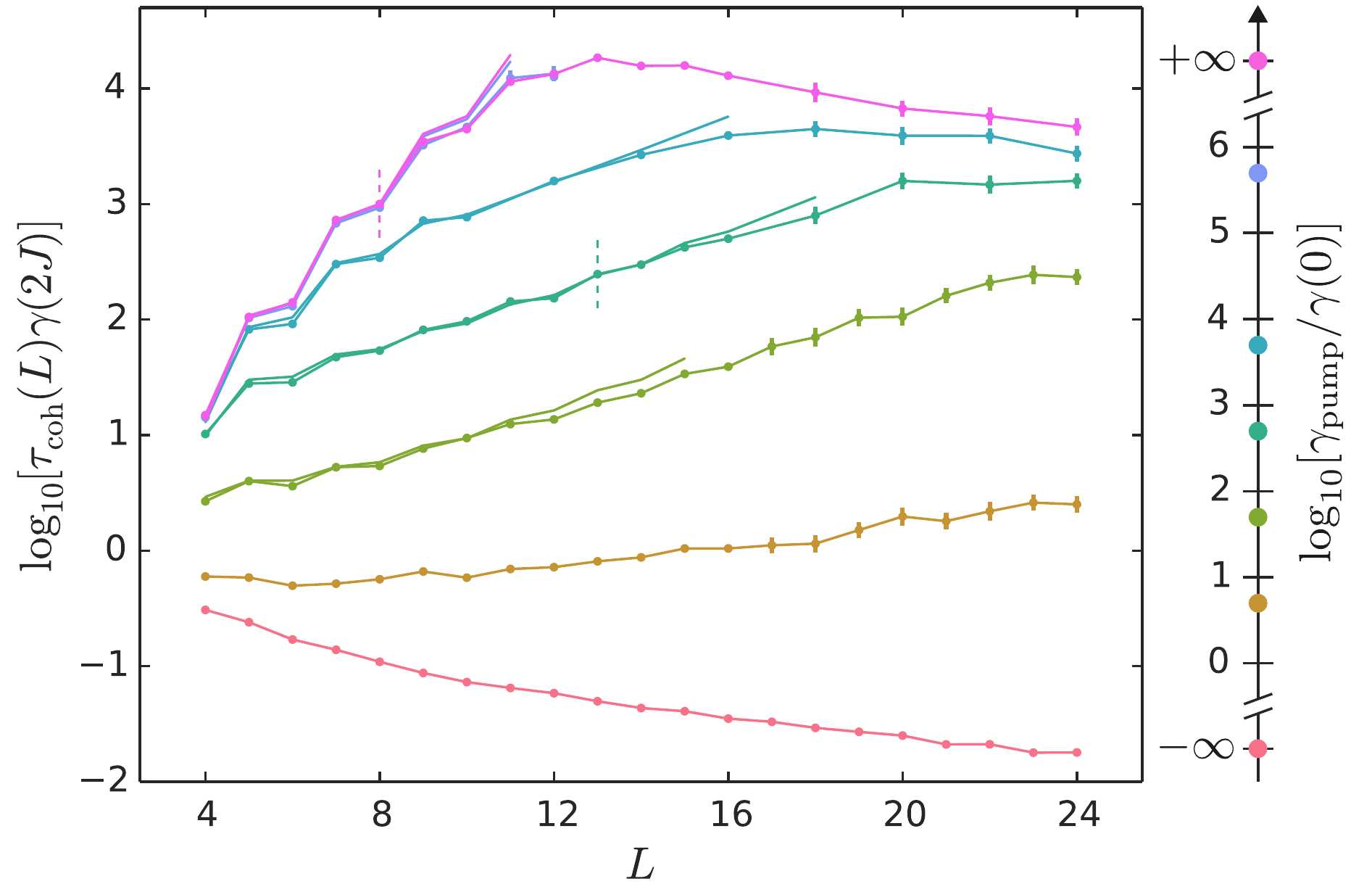}
	\caption{Coherence time as a function of system size, for different potential-pump rates $\gamma_\mathrm{pump}$ and fixed $U/T = 5$. The potential-decay rate is set to $\gamma_\mathrm{dec} = 10^{-3}$, except for the uppermost curve where it is chosen 10 times larger (note that the two upper curves nevertheless match, and that they coincide with the second curve from the top in Fig.~\ref{fig:coherence_time_vs_U}). Data points and error bars are obtained by a kinetic Monte Carlo simulation of the full system dynamics with $10^3$ trajectories. The upper five curves are fitted using a single function of the form $a P^{(0)}_\mathrm{sep}(\ell) + b P_\mathrm{esc} c^{\ell-2}$, with parameters $(a,b,c,U_\mathrm{fit}/T) \approx (0.84, 3.24, 0.40, 4.82)$. The values $(a,U_\mathrm{fit}/T)$ are extracted by fitting the uppermost curve using points between $L = 4$ and the dashed vertical bar. The remaining parameters are extracted in a similar way using the fourth curve from the top. The resulting curves, shown up to the point where deviations occur, are in remarkable agreement with our theoretical model [Eq.~\eqref{eq:1DResultFinitePump}]. As in Fig.~\ref{fig:coherence_time_vs_U}, crossovers are consistent with $L_\mathrm{max} \approx 13.7$. The fact that saturation occurs at larger sizes for decreasing pump rate is consistent with the fact that reducing $\gamma_\mathrm{pump}$ effectively lowers the maximum potential depth $U$, which increases $L_\mathrm{max}$.}
	\label{fig:coherence_time_vs_pump}
\end{figure}

In Fig.~\ref{fig:coherence_time_vs_decay}, we investigate in more detail the robustness of our scheme with respect to varying the trench-decay rate $\gamma_\mathrm{dec}$. Our numerical results clearly demonstrate that our scheme remains effective up to system sizes $L \approx L_\mathrm{max}$ for values $\gamma_\mathrm{dec}$ in a wide parameter range around the ``optimal'' value identified in Eq.~\eqref{eq:optDecayRate}. These results indicate that trench saturation imposes a hard constraint $L \lesssim L_\mathrm{max}$ on our scheme, while trench refilling and splitting lead to much softer requirements on $\gamma_\mathrm{dec}$. More specifically, error correction is still effective when the upper and lower bounds of Eq.~\eqref{eq:boundsOnGammaDec} are not strictly satisfied. We remark that the fact that a large range of values $\gamma_\mathrm{dec}$ is suitable for our scheme is expected for system sizes $L < L_\mathrm{max}$. Indeed, the left-hand-side of Eq.~\eqref{eq:boundsOnGammaDec} shows that trench refilling should only be relevant when $\gamma_\mathrm{dec} \lesssim (T/U)\gamma(2J-2U)(L/2)^{\eta-\eta'}$, which scales exponentially with temperature (as $\mathrm{e}^{-2(J-U)/T}$). In contrast, the right-hand-side of Eq.~\eqref{eq:boundsOnGammaDec}, which describes trench splitting, requires $\gamma_\mathrm{dec} \lesssim 2 \eta (T/U) \gamma(0)(L/2)^{-2}$, which scales quadratically with $T$. For system sizes $L < L_\mathrm{max}$, the range of suitable values for $\gamma_\mathrm{dec}$ is therefore exponentially large. Note that we recover a similar scaling as for the usual toric code when the trench-decay rate is so small that trench refilling strongly dominates (bottom curve in Fig.~\ref{fig:coherence_time_vs_decay}). Due to the slow decay, the entire system ends up being covered by a single trench, at which point the situation reduces to the standard toric code with $2J-2U$ as the relevant gap (instead of $2J$).

In Fig.~\ref{fig:coherence_time_vs_pump}, we verify our theoretical prediction that the exponential increase of the memory-coherence time survives when the potential-pump rate $\gamma_\mathrm{pump}$ is finite. As long as $\gamma_\mathrm{pump}$ is significantly larger than the diffusion rate $\gamma(0)$, we find that $\tau_\mathrm{coh}$ simply follows a slower exponential increase, as expected from our theoretical model [Eq.~\eqref{eq:1DResultFinitePump}]. As in Fig.~\ref{fig:coherence_time_vs_U}, the observed crossovers are consistent with our theoretical estimates for $L_\mathrm{max}$. Saturation takes place at larger system sizes for lower potential-pump rates, which stems from the fact that decreasing $\gamma_\mathrm{pump}$ effectively reduces the depth of potential trenches, thereby increasing $L_\mathrm{max}$ [see discussion below Eq.~\eqref{eq:maxCoherenceTime}].

Our analysis demonstrates that the coherence-time improvement provided by our scheme does not crucially depend on the potential-decay rate. Provided that $L \lesssim L_\mathrm{max}^\mathrm{sat}$ and Eq.~\eqref{eq:boundsOnGammaDec} is approximately satisfied, the quantitative increase of the memory time with system size is determined by two more crucial quantities: First, the maximum potential depth $U$, which has a striking effect on the coherence time even for moderate values $U/T \gtrsim 1$. Second, the potential-pump rate, which should be much larger than the diffusion rate to fully benefit from having a large $U$. We remark that the parameter regimes explored in our simulations provide a memory lifetime $3-4$ orders of magnitude longer than the inverse local-error rate $1/\gamma(2J)$, in stark contrast to the usual toric code (bottom curve in Fig.~\ref{fig:coherence_time_vs_U}). We expect to obtain much stronger enhancements in ideal regimes which cannot be explored numerically.

\section{Conclusion}
\label{sec:conclusion}

Protecting quantum bits against environmental errors remains one of the outstanding challenges towards practical quantum computing. In this work, we have proposed a passive and efficient way to correct such errors in the context of topological quantum memories based on stabilizer codes, such as the toric code. Our scheme relies on driven-dissipative ancilla systems that couple to the memory and make elementary excitations (anyons) dig their own potential ``grave''. When anyons are created, they rapidly form a potential trench in which they get trapped, which strongly suppresses anyon-pair separation over anyon recombination. The required ancilla systems act in a local, translation-invariant way, and are simple enough to potentially lend themselves to practical implementations. We have outlined a potential realization based on circuit-QED systems.

Our theoretical analysis and extensive numerical simulations have revealed three important features: First, the probability that anyons created as a pair separate by a distance $\ell$ decreases exponentially with $\ell$. This scaling is observed up to a maximum separation $\ell_\mathrm{max}$ corresponding to the typical length scale at which potential trenches split due to their finite decay rate. Second, the memory-coherence time increases exponentially with system size, up to an upper bound $L_\mathrm{max}$ beyond which limiting effects (trench saturation, refilling and splitting) come into play. Although we have derived an optimal decay rate $\gamma_\mathrm{dec}$ from theoretical estimates, our simulations demonstrate that the coherence time increases exponentially up to $L_\mathrm{max}$ in a wide parameter range of $\gamma_\mathrm{dec}$. Third, and most importantly, the memory lifetime improves up to double exponentially with inverse temperature, in stark contrast with the Arrhenius law applicable to generic qubits.

Pump and decay (drive and dissipation) are key ingredients of our scheme. The potential-pump rate, which controls how fast anyons generate their self-trapping potential, determines the rate at which ancilla systems acquire information about the memory. It must be larger than the typical diffusion rate of anyons for our scheme to be effective. The potential-decay rate also plays an crucial role, to ``erase'' potential trenches when anyon pairs recombine. More generally, it is the driven-dissipative nature of the ancilla systems which allows to remove entropy from the memory.

Several extensions of our scheme will be interesting to explore: In this work, we have assumed that perturbations caused by the environment are strictly local. To protect the memory against quasi-local errors, one could consider ancilla systems that force anyons to generate more extended potential trenches. This could also help reduce the required strength of the potential-pump rate. More generally, our scheme could prove useful, with modifications, as a dynamical decoder~\cite{Herold2015,Herold2015_2}. In this context, one could imagine to decode the memory using classical cellular automata that perform simple sequences of local updates (of the anyon positions and local potentials), thereby reproducing the dynamics described in this work and gradually correcting errors over time.

Further study will be required to construct specific implementations of our scheme. Beyond analog simulation using, e.g., circuit-QED systems, digital (or gate-based) quantum simulation could be considered~\cite{Muller2011}. More generally, our error-correction mechanism can be readily applied to generalizations of the toric code such as the surface code~\cite{Fowler2012}. A first step would be to consider implementations based on the repetition code, its 1D building block. This 1D code was recently realized in circuit-QED systems, to demonstrate active error correction~\cite{Kelly2015}. We except our scheme to allow for the demonstration of efficient passive error correction in similar settings.

\section*{Acknowledgements}

We thank Gil Refael for his support and feedback. We are also grateful to John Preskill, Spyridon Michalakis, and Fernando Pastawski for valuable discussions. This work was funded by the Swiss National Science Foundation (SNSF) and the Institute for Quantum Information and Matter, an NSF Physics Frontiers Center with support of the Gordon and Betty Moore Foundation (Grant No. GBMF1250). C.-E. B. acknowledges support from the National Science Foundation (NSF) under Grant No. DMR-1410435 and PHY11-25915.

\appendix

\section{Implementation of the ancilla systems}
\label{app:ancillaSystems}

In this section, we outline a potential implementation of our scheme based on cavity quantum electrodynamics (cavity QED), which we consider as a promising platform for our proposal. We discuss how to realize the driven-dissipative ancilla systems that are key to our scheme, and how the latter are coupled to the stabilizers of the quantum memory. As in the main text, we focus on the implementation of the ancilla systems coupled to the plaquette operators $B_p = \prod_{j \in p} \sigma_j^z$ of the toric code. A similar construction can be envisioned for star operators $A_s = \prod_{j \in s} \sigma_j^x$ [Eq.~\eqref{eq:stabilizers}].

We start by recalling the effective plaquette-ancilla coupling Hamiltonian [Eq.~\eqref{eq:effectiveHamAncilla}] which we would like to achieve:
\begin{equation}
    H_{p, \mathrm{eff}}(t) = \frac{U}{2} n_\mathrm{c}(t) B_p,
    \label{eq:effPlaqAncillaHam}
\end{equation}
where $n_\mathrm{c}(t)$ denotes the time-dependent occupation of some ancilla level. In what follows, we consider an ancilla population that consists of photons in a particular cavity mode with frequency $\omega_\mathrm{c}$ and creation (annihilation) operator $a^\dagger_\mathrm{c}$ ($a_\mathrm{c}$), respectively, such that $n_\mathrm{c}(t) = \langle a^\dagger_\mathrm{c} a_\mathrm{c} \rangle(t)$ (expectation value in the ancilla state). To satisfy the requirements of our scheme, the occupation of this ancilla-cavity mode should (i) be pumped at a fast rate $\gamma_\mathrm{pump}$ (to a maximum value which we set to $1$, for simplicity) when the plaquette is occupied ($B_p = -1$), and (ii) should decay at a slow rate $\gamma_\mathrm{dec}$ when the plaquette is empty ($B_p = +1$). To implement this dynamics, we first consider the more concrete plaquette-ancilla coupling
\begin{equation}
    H_{p-a} = \frac{U}{2} a^\dagger_\mathrm{c} a_\mathrm{c} \prod_{j \in p} \sigma_j^z,
    \label{eq:dispPlaqAncillaCoupling}
\end{equation}
which takes a natural form judging by Eq.~\eqref{eq:effPlaqAncillaHam}. This Hamiltonian describes a \emph{dispersive} coupling between the plaquette operator and the ancilla-cavity mode; it involves no exchange of excitations between the system and the ancilla. Instead, it tells us that the occupation of the cavity mode shifts the plaquette energy, and vice versa.

The dispersive coupling described by Eq.~\eqref{eq:dispPlaqAncillaCoupling} makes it straightforward to pump photons into the ancilla cavity conditioned on the plaquette occupation, as desired. To achieve this, one can simply introduce a coherent driving field (laser) with frequency $\omega_\mathrm{d}$ tuned to the shifted frequency $\omega_\mathrm{c} + U/2$ of the cavity mode found when the plaquette is occupied. The dispersive shift $U$ should be much larger than the amplitude $\Omega_\mathrm{d}$ of the driving field (it is also typically smaller than $\omega_\mathrm{c}$). Strong dispersive shifts have been observed, e.g., in cavity QED-systems based on superconducting qubits (circuit QED)~\cite{Schuster2007}.

To ensure that the ancilla-cavity occupation does not rise above a specific number (which we choose as $1$, here), we assume that the ancilla cavity is \emph{nonlinear}, with a nonlinearity $U_\mathrm{c}$ much larger than the drive amplitude (but smaller than $\omega_\mathrm{c}$). As desired, this nonlinearity forbids the introduction of more than one photon in the cavity when driving the latter at a frequency $\omega_\mathrm{d} \approx \omega_\mathrm{c} + U/2$. In this photon-blockade regime, the maximum number of photons that can be pumped into the cavity is one (up to perturbative corrections which are irrelevant here).

Although the direct coherent-pumping scheme described so far satisfies most of our requirements, it also leads to one complication: It makes the cavity mode occupation oscillate in time (Rabi oscillations) with frequency corresponding to the drive amplitude (on resonance). To suppress coherent effects from the drive, one can consider an incoherent pump consisting, e.g., of a coherent tone with frequency $\omega_\mathrm{d} \sim \omega_\mathrm{c} + U/2$ modulated by some finite-bandwidth noise (see, e.g., Ref.~\cite{Hoffman2011}). A more elegant way to achieve this was considered in Ref.~\cite{Kapit2014} (for different purposes) using an auxiliary ancilla qubit with a \emph{fast} decay rate $\gamma_\mathrm{q} \sim \Omega_\mathrm{d} \gg \gamma_\mathrm{c}$, where $\gamma_\mathrm{c}$ is the intrinsic photon-loss rate of the ancilla cavity (and $\Omega_\mathrm{d}$ the amplitude of the drive). By driving this ancilla qubit \emph{together} with the ancilla-cavity mode using a two-photon \emph{parametric} drive, one obtains an effective pumping rate $\gamma_\mathrm{pump} \sim \gamma_\mathrm{q}$ for the ancilla-cavity mode. The latter can be much larger than the cavity decay rate $\gamma_\mathrm{dec} \sim \gamma_\mathrm{c}$, as required for the error-correcting scheme presented in the main text. Intuitively, this comes from the fact that the parametric drive can only add (remove) photons with frequency $\omega_\mathrm{c} + U/2$ to the ancilla cavity \emph{together} with exciting (de-exciting) the auxiliary qubit. Since the fast qubit decay ensures that the qubit spends most of its time in its ground state, the cavity can only (mostly) be replenished, at a fast rate controlled by the qubit decay. We refer the reader to Ref.~\cite{Kapit2014} for details regarding the physical implementation of such a scheme in circuit QED. Note that a large separation of scales $\gamma_\mathrm{pump} \gg \gamma_\mathrm{dec}$ can be achieved in this setting (see, e.g., Refs.~\cite{Houck2012,Schmidt2013}).

We remark that the Hamiltonian of Eq.~\eqref{eq:dispPlaqAncillaCoupling} corresponds to a five-body interaction, which may be challenging to realize in practice. One could imagine, however, to generate a similar dynamics using two-body interactions
\begin{equation}
    H_{p-a} = \frac{U}{2} a^\dagger_\mathrm{c} a_\mathrm{c} \sum_{j \in p} \sigma_j^z,
    \label{eq:dispPlaqAncillaCouplingSum}
\end{equation}
where we have replaced the product in Eq.~\eqref{eq:dispPlaqAncillaCoupling} by a sum. To understand why this could also be suitable for our scheme, let us first note that both operators $\prod_{j \in p} \sigma_j^z$ and $\sum_{j \in p} \sigma_j^z$ share the same eigenstates. The main difference is that the sum operator can have five different eigenvalues $(0, \pm 2, \pm 4)$ instead of two for the product (plaquette) operator $(\pm 1)$. Consequently, the coupling of the ancilla-cavity mode will lead to a splitting into five cavity modes with frequency centered around the bare cavity frequency $\omega_\mathrm{c}$, instead of two. Among those, only two of these eigenvalues ($\pm 2$) correspond to eigenstates of the plaquette operator with $\prod_{j \in p} \sigma_j^z = -1$, i.e., where the plaquette is occupied. Therefore, instead of pumping the ancilla cavity mode at a single frequency, one could achieve the desired pumping conditioned on the plaquette occupation by using two pumping schemes with frequency $\omega_\mathrm{c} + U$ and $\omega_\mathrm{c} - U$. The effective plaquette-ancilla Hamiltonian would take the form of Eq.~\eqref{eq:effPlaqAncillaHam} with $U/2 \to U$.

The main difference between implementations based on Hamiltonians~\eqref{eq:dispPlaqAncillaCoupling} and~\eqref{eq:dispPlaqAncillaCouplingSum} comes from dephasing. Indeed, both types of dispersive couplings not only effectively shift the plaquette energy, as desired, but also lead to dephasing due fluctuations of the ancilla-cavity-mode occupation induced by driving and damping~\cite{Gambetta2006,Clerk2010}. Although dephasing of the plaquette operator $\prod_{j \in p} \sigma_j^z$ is irrelevant, dephasing at the level of individual spin operators $\sigma_j^z$ would generate errors from the viewpoint of star operators $\sim \prod_{j \in s} \sigma_j^x$, and should therefore be avoided if one wants to correct both bit-flip and phase-flip logical errors. Ultimately, dephasing must be compared to the gap of the toric code, which we assume to be much larger. Finally, we note that, although dispersive couplings of the form~\eqref{eq:dispPlaqAncillaCouplingSum} are routinely achieved in circuit-QED systems, achieving the analog with $\sigma_j^x$ operators (as necessary for star operators) could be more challenging. Specific realizations of our scheme will be examined in future work.


\section{Details of the 1D model}
\label{app:details1DModel}

\subsection{Details of the derivation of the pair-separation probability}
\label{app:derivationPsep}

\begin{figure*}[t]
    \includegraphics[width=0.93\textwidth]{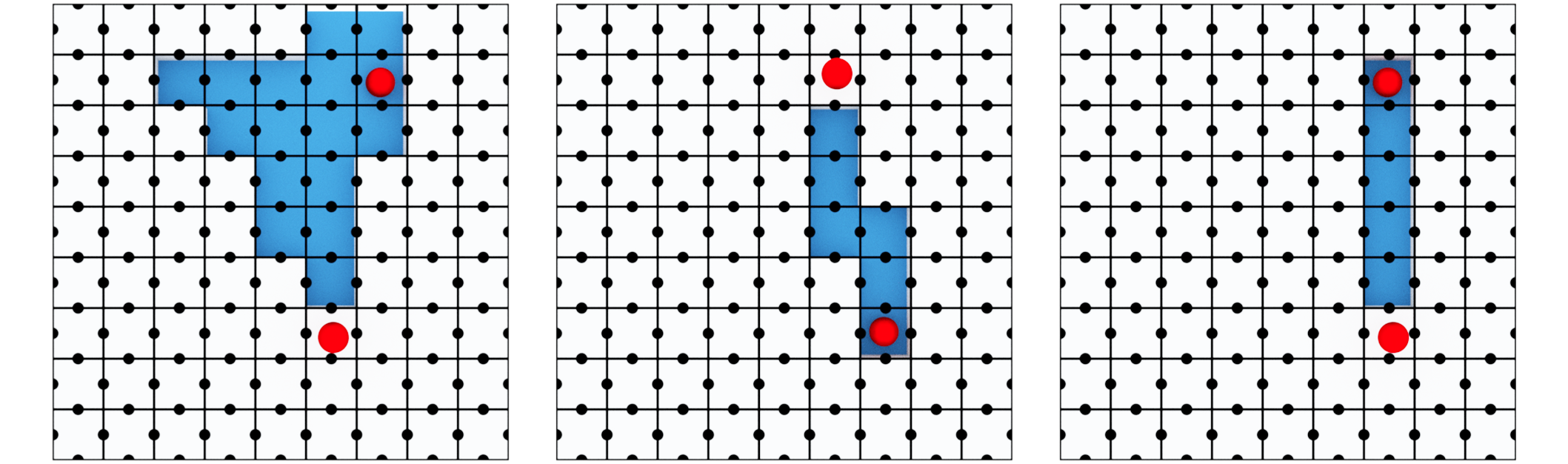}
    \caption{Typical shape of error-causing potential trenches for increasing maximum potential depth $U = 0.2$, $0.6$, $0.8$ (from left to right) and fixed bath temperature $T = 0.15$. The above snapshots are taken during a kinetic Monte Carlo simulation of the anyon-pair dynamics (see Appendix~\ref{app:simulations}) at the time when a logical error first occurs, i.e., when one of the anyons manages to overcome the energy barrier defined by the trench and separates from its partner by more than half the system size $L = 9$ (here, in the $y$ direction). The shape of the relevant trenches changes from two-dimensional at low $U \sim T$ to one-dimensional for $U \gg T$. The observed configurations are in good qualitative agreement with the estimated crossover lengths $\ell_{1\mathrm{D}}(U) \approx 2.1$, $3.9$, $5.6$. The rate $\gamma(2J)$ for anyon-pair creation is artificially set to zero after one pair is created. Other relevant parameters are chosen as $J = 2$, $\kappa = 1$, with infinite potential-pump rate $\gamma_\mathrm{pump}$ and vanishing decay rate $\gamma_\mathrm{dec}$.}
    \label{fig:typical_trenches}
\end{figure*}

In this section, we derive an explicit expression for the quantity $P_\mathrm{rec}(n)$ used in deriving Eq.~\eqref{eq:Pl/Pl-1}, which represents the probability that an anyon located at $x = n$ in a 1D trench recombines with its partner assumed to be fixed at $x = 0$ without ever going to $x = n+1$. To solve this 1D free-diffusion process (we assume that the anyon remains in the 1D trench), we first derive the probability $P_{l}(r)$ that a random walker in 1D eventually takes $r$ steps to the right without ever taking $l$ steps to the left (with respect to its initial position). If diffusion is symmetric, this probability is invariant under switching ``right'' ($+1$) and ``left'' ($-1$) directions. Since the probability of eventually taking $r$ steps to the right or $l$ steps to the left is equal to $1$, we can write
\begin{equation}
    P_{l}(r) + P_{r}(l) = 1.
    \label{eq:free_diff_1}
\end{equation}
We also find
\begin{eqnarray}
    P_{1}(1) & = & \frac{1}{2} \\
    P_{1}(2) & = & P_{1}(1) P_{2}(1) \\
    P_{1}(n) & = & \prod_{i=1}^{n} P_{i}(1) = \prod_{i=1}^{n}[1-P_{1}(i)],
\end{eqnarray}
which yields the recurrence relation
\begin{equation}
    P_{1}(n) = \frac{P_1(n-1)}{1+P_1(n-1)},
\end{equation}
with solution
\begin{equation}
    P_{1}(n) = \frac{1}{n+1}.
    \label{eq:free_diff_2}
\end{equation}
As mentioned in the main text, the probability $P_\mathrm{rec}(n)$ only differs from $P_1(n)$ because of the enhanced probability to go from $x = 1$ to $x = 0$ in the last recombination step, since the relative rate between anyon recombination and anyon diffusion is $\alpha_\downarrow > 1$. The modification due to this last step yields
\begin{eqnarray}
    \frac{P_\mathrm{rec}(n)}{P_1(n-1)} & = & \frac{\alpha_\downarrow}{1+\alpha_\downarrow} + \frac{1}{1+\alpha_\downarrow}P_{n-1}(1)\frac{P_\mathrm{rec}(n)}{P_1(n-1)} \nonumber \\
    \Rightarrow P_\mathrm{rec}(n) & = & \frac{\alpha_\downarrow P_1(n-1)}{1+\alpha_\downarrow-P_{n-1}(1)}.
\end{eqnarray}
Using Eqs.~\eqref{eq:free_diff_1} and~\eqref{eq:free_diff_2}, we thus obtain the expression given in the main text:
\begin{equation}
    P_\mathrm{rec}(n)=\frac{1}{n+\alpha_\downarrow^{-1}}.
\end{equation}

\subsection{Regime of applicability of the 1D model}
\label{app:applicability1D}

To identify the expected regime of applicability of our 1D toy model, it is useful to examine the typical shape of potential trenches. In the limit of a maximum potential depth $U \ll T$, it is clear that anyons essentially diffuse according to a (symmetric) 2D random walk as in past studies of the toric code coupled to a thermal bath~\cite{Brown2014_2}. Potential trenches only become relevant when $U$ is comparable or larger to the bath temperature $T$~\footnote{We define potential trenches as domains of connected plaquettes identified by $U > T$.}. For finite $U \gtrsim T$, the typical trenches observed when a memory error occurs are mainly two-dimensional (see Fig.~\ref{fig:typical_trenches}). For $U \gg T$, however, they become one-dimensional. Intuitively, this can be understood by noticing that the probability of extending a trench by one plaquette in any direction ($\propto \alpha_\uparrow$) is exponentially suppressed in $U/T$ when $U \gg T$. Consequently, the most likely way to cause a logical error is to extend a trench a minimum number of times, which naturally leads to 1D shapes. The trench orientation is fixed by the shortest error-causing path, which depends on how logical operators are defined.

The above argument suggests that our 1D ``toy'' model provides a suitable description of the error probability for $T \ll U$, which coincides with the regime where potential trenches are relevant~\footnote{Our 1D model readily applies, for any temperature $T$, to implementations of our scheme based on 1D stabilizer codes (see, e.g., Ref.~\cite{Pedrocchi2015}).}. To quantify this, one can consider an anyon pair in a 1D trench and introduce the possibility of lateral motion in a perturbative way. Specifically, one can consider diffusion along the trench where the probability that an anyon extends the trench and ultimately cause an error is reduced due to the possibility to ``escape'' in the lateral direction. As we demonstrate in the next section, this decreases the probability of each extension step by a factor $(1 - 2 l^2 \alpha_\uparrow/3)$. When the accumulated suppression of probability to extend the trench to a size $\ell$ becomes of order $1$, i.e., when $\ell \approx [9/(2 \alpha_\uparrow)]^{1/3}$, a substantial amount of error-causing trenches is expected to be two-dimensional. This yields the length scale presented in Eq.~\eqref{eq:ell1D} of the main text.

\subsection{Corrections in the quasi-1D regime}
\label{app:quasi1DCorrections}

The pair-separation probability $P^{(0)}_\mathrm{sep}(\ell)$ used in the analysis presented in the main text [see Eq.~\eqref{eq:1DResult}] is derived under the assumption that anyons can only move \emph{along} the 1D trench that they generate. In what follows, we derive the analog of this quantity in the quasi-1D regime where the potential trench is one-dimensional but anyons can nevertheless escape in the \emph{lateral} direction. To ensure that the trench remains one-dimensional, we treat such escape events as ``losses'', as clarified below. Corrections due to the finite probability of lateral motion allow us to estimate the limits of applicability of the 1D model used in the main text. For clarity, we adopt similar notations as in the main text and indicate quantities pertaining to the quasi-1D regime by an asterisk (``$^*$'').

In the absence of lateral motion, the pair-separation probability (which we call here ``trench-extension'' probability) satisfies the recurrence relation given in Eq.~\eqref{eq:Pl/Pl-1} of the main text. Including the possibility of lateral losses leads to the following modifications:
\begin{align}
     \ \ \ \ \frac{P^*_\mathrm{ext}(n)}{P^*_\mathrm{ext}(n-1)} = & \frac{\alpha_\uparrow}{1+3\alpha_\uparrow} \nonumber \\
    + \frac{1}{1+3\alpha_\uparrow} & \left[\bar{P}_\mathrm{rec}^*(n-1)\frac{P_\mathrm{ext}^*(n)}{P^*_\mathrm{ext}(n-1)}\right].
    \label{eq:P_n_n-1_2D}
\end{align}
Here, differences come from the fact that free diffusion inside the trench is modified due to losses in the lateral direction, such that some of the relations used in Sec.~\ref{app:derivationPsep} no longer hold (Eq.~\eqref{eq:free_diff_1}, in particular). In that case, the probability $\bar{P}_\mathrm{rec}^*(n)$ to take one step to the right without recombining with the other anyon (located a distance $n$ to the left) is also no longer given by $1-P_\mathrm{rec}^*(n)$. Instead, it satisfies the recurrence relation
\begin{equation}
    \bar{P}_\mathrm{rec}^*(n) =  P^*_{n-1}(1) + P^*_{1}(n-1){P}_\mathrm{1,\alpha_\downarrow}^*(n)\,,
    \label{eq:quasi1D_1}
\end{equation}
where $P_\mathrm{1,\alpha_\downarrow}^*(n)$ is the analog of the probability $P_1^*(n)$ in the situation where the recombination step that should be avoided (from $x = 1$ to $x = 0$) occurs with a relative rate $\alpha_\downarrow$ as compared to diffusion. We find that $P_\mathrm{1,\alpha_\downarrow}^*(n)$ takes a simple form
\begin{equation}
    P_{1,\alpha_\downarrow}^*(n) = \frac{P^*_{1}(n-1)}{1+\alpha_\downarrow+2\alpha_\uparrow-P^*_{n-1}(1)},
    \label{eq:quasi1D_2}
\end{equation}
which follows from the recurrence relation
\begin{eqnarray}
    P^*_{1,\alpha_\downarrow}(n) & = & \frac{1}{1+\alpha_\downarrow+2\alpha_\uparrow}P^*_{2,\alpha_\downarrow}(n-1) \nonumber \\
    & = & \frac{P^*_{1}(n-1)+P^*_{n-1}(1)P^*_{1,\alpha_\downarrow}(n)}{1+\alpha_\downarrow+2\alpha_\uparrow}.\ \ \ \ \ \ \ 
\end{eqnarray}
The remaining ingredients are the two probabilities $P^*_{n}(1)$ and $P^*_{1}(n)$ for free diffusion in the quasi-1D regime. As we show in Sec.~\ref{app:free_diffusion_quasi1D}, the results in Eqs.~\eqref{eq:free_diff_2} and~\eqref{eq:free_diff_1} become, in the absence of lateral motion,
\begin{eqnarray}
    P^*_{1}(n) & = & \frac{1}{n+1}\left(1-\frac{n(n+2)}{3}\alpha_\uparrow\right) \label{eq:P1n_expand} \\
    P^*_{n}(1) & = & \frac{n}{n+1}\left(1-\frac{2n+1}{3}\alpha_\uparrow\right), \label{eq:Pn1_expand}
\end{eqnarray}
to leading order in the relative rate $\alpha_\uparrow$ between escaping the trench laterally (at each step) and diffusing along the latter. Note that both probabilities are reduced because of ``losses'' in the lateral direction. Combining these results with Eqs.~\eqref{eq:quasi1D_1} and~\eqref{eq:quasi1D_2}, we obtain
\begin{align}
    \bar{P}^*_{\rm rec}(n) & = 1 - \frac{1}{n+\alpha_\downarrow^{-1}} \nonumber \\
    & - n\alpha_\uparrow \frac{[6+\alpha_\downarrow(n-1)][6+\alpha_\downarrow(2n-1)]}{3(1+\alpha_\downarrow n)^{3}},
\end{align}
(to leading order in $\alpha_\uparrow$), which finally yields
\begin{eqnarray}
    P^*_\mathrm{ext}(n) & = & \alpha_\uparrow^{n} \prod_{k=1}^{n}\bigg[n-1+\alpha_\downarrow^{-1} -\alpha_\uparrow \bigg(\frac{2}{3}n^{3}+2\alpha_\downarrow^{-1}n^{2} \nonumber \\
    & & +2\alpha_\downarrow^{-2}n-\frac{5}{3}n+\alpha_\downarrow^{-2}-2\alpha_\downarrow^{-1}+1\bigg)\bigg]. \ \ \ \ \
\end{eqnarray}
In the limit $n \gg \alpha_\downarrow$, we find the simplified expression
\begin{equation}
    P^*_\mathrm{ext}(n) = \alpha_\uparrow^{n} \prod_{k=1}^{n}\left[\left(n-1+\alpha_\downarrow^{-1}\right)\left(1-\frac{2}{3}n^{2}\alpha_\uparrow\right)\right],
\end{equation}
which provides the correction factor $1 - 2 n^2 \alpha_\uparrow/3$ used in the main text.

\subsubsection{Further details of the derivation}
\label{app:free_diffusion_quasi1D}

The probabilities $P^*_{n}(1)$ and $P^*_{1}(n)$ corresponding to free diffusion in the quasi-1D regime can be determined recursively using the following relations:
\begin{eqnarray}
    P^*_{1}(n) & = & P^*_{1}(n-1)P^*_{n}(1) \label{eq:P1nPn1} \\
    P^*_{n}(1) & = & P^*_{n-1}(1) + P^*_{1}(n-1)P^*_{1}(n) \label{eq:Pn1P1n} \\
    P^*_{n}(1) & = & \frac{1}{2+2\alpha_\uparrow} + \frac{1}{2+2\alpha_\uparrow}P^*_{n-1}(1)P^*_{n}(1).\;
    \label{eq:Pn1_recursive}
\end{eqnarray}
Plugging Eq.~\eqref{eq:P1nPn1} in \eqref{eq:Pn1P1n} yields
\begin{equation}
    P^*_{1}(n) = \sqrt{P^*_{n}(1)\left(P^*_{n}(1)-P^*_{n-1}(1)\right)},
\end{equation}
which allows us to determine $P^*_{1}(n)$ given $P^*_{n}(1)$. The latter can be found recursively from Eq.~\eqref{eq:Pn1_recursive},
\begin{equation}
    P^*_{n}(1) = 1 - \frac{1+2\alpha_\uparrow-P^*_{n-1}(1)}{2+2\alpha_\uparrow-P^*_{n-1}(1)}.
\end{equation}
We now define $P^*_{n}(1) \equiv 1 - h_{n}/k_{n}$ with initial conditions $h_{0} = 1$, $k_{0} = 1$ (below we also need $h_{1} = 1 + 2\alpha_\uparrow$). The recurrence relations for $h_{n}$ and $k_{n}$ become
\begin{eqnarray}
    h_{n} & = & 2\alpha_\uparrow k_{n-1} + h_{n-1} \\
    k_{n} & = & (1+2\alpha_\uparrow)k_{n-1} + h_{n-1}.
\end{eqnarray}
Plugging the first equation into the second leads to
\begin{equation}
    h_{n+1} = (2+2\alpha_\uparrow)h_{n}-h_{n-1},
\end{equation}
which can be solved by a standard ansatz of the form $h_{n} = r^{n}$. The latter leads to a characteristic quadratic equation for $r$ with solutions $r_{1/2} = 1 + \alpha_\uparrow \pm \sqrt{(2+\alpha_\uparrow)\alpha_\uparrow}$. The full solution then takes the form $h_{n} = ar_{1}^{n} + br_{2}^{n}$, with coefficients $a,b$ set by initial conditions. We find
\begin{eqnarray}
    h_{n} & = & \frac{r_{1}^{n}}{1+r_{2}}+\frac{r_{2}^{n}}{1+r_{1}} \\
    k_{n} & = & \frac{1}{2\alpha}\left(\frac{r_{1}-1}{1+r_{2}}r_{1}^{n}+\frac{r_{2}-1}{1+r_{1}}r_{2}^{n}\right),
\end{eqnarray}
which leads to a rather lengthy expression for $P^*_{n}(1)$ {[}and subsequently for $P^*_{1}(n)${]}. The results, expanded to linear order in $\alpha_\uparrow$, are given in Eqs.~\eqref{eq:Pn1_expand} and~\eqref{eq:P1n_expand}.

\section{Trench percolation}
\label{app:trenchPercolation}

Trench percolation occurs when multiple trenches of length $\ell < L/2$ combine to form a long error-causing trench of length $\sim L/2$. Small trenches are more likely to percolate than large ones. To understand why, let us estimate the probability $P_n(\ell)$ of finding at least $n$ trenches of similar length $\ell < L/2$ within the lifetime $\tau_\mathrm{t}(\ell)$ of a single trench. The rate at which such trenches are created is given by $\gamma_\mathrm{t}(\ell) \approx 2 L^2 \gamma(2J) P_\mathrm{sep}(\ell)$ [see Eq.~\eqref{eq:tau_cre}], which decreases rapidly with increasing $\ell$ due to the exponential or large power-law suppression of $P_\mathrm{sep}(\ell)$ with $\ell$ [see Eq.~\eqref{eq:1DResultFinitePump}]. In comparison, the lifetime of a single trench, $\tau_\mathrm{t}(\ell) \approx (\ell/2)^2/\gamma(0) + \gamma_\mathrm{dec}^{-1}$~\footnote{The first term corresponds to the typical time required for anyons to diffuse and recombine. The second corresponds to the subsequent trench decay.}, scales very weakly with $\ell$, such that $\mu_\mathrm{t}(\ell) \equiv \gamma_\mathrm{t}(\ell) \tau_\mathrm{t}(\ell) < 1$ is typically satisfied for large trenches. Assuming that trenches are created independently (which is a reasonable assumption for low trench densities), Poisson statistics dictates that the probability of finding at least $n$ trenches in the time interval $\tau_\mathrm{t}(\ell)$ is $P_n(\ell) = \sum_{m \geq n} \mathrm{e}^{-\mu_\mathrm{t}(\ell)} \mu_\mathrm{t}(\ell)^m/m!$. In general, $P_n(\ell)$ decreases faster than exponentially with $n$ in the regime $n \gg \mu_\mathrm{t}(\ell)$. While this is naturally satisfied for large trenches with $\mu_\mathrm{t}(\ell) \ll 1$, it need not be so for smaller trenches with a significant value $\mu_\mathrm{t}(\ell)$. Since the number $n$ of trenches required for percolation scales with the area $L^2$ of the system, i.e., $n \sim (L/\ell)^2$, percolation can be suppressed by enforcing $\mu_\mathrm{t}(\ell) \ll (L/\ell)^2$ for any $\ell$ (for $\ell \sim \mathcal{O}(1)$, in particular). This leads to the following approximate lower bound for the trench decay rate:
\begin{equation}
    \gamma_\mathrm{dec} \gg \gamma(2J).
    \label{eq:gammaDecLowerBound2}
\end{equation}
This condition tells us that percolation is suppressed provided that the trench decay rate is much larger than the rate at which new trenches (or anyon pairs) are created. This lower bound, however, is less constraining than the one obtained for trench refilling [Eq.~\eqref{eq:gammaDecLowerBound1}].

\section{Simulations}
\label{app:simulations}

Within our model, the dynamics of the coupled toric-code-ancilla system is generated by local spin flips $\sim \sigma_j^x$ induced by the thermal bosonic bath. These flips are, by assumption, uncorrelated. More importantly, they cause transitions between anyon configurations at a rate that depends on time, i.e., on the current anyon configuration as well as on the current values of the effective plaquette potentials induced by the ancilla systems. The resulting dynamics is captured by the rate equation~\eqref{eq:rateEquation}, with rates $\gamma(\omega_{mn})$ that depend on time through the time-dependent energy difference $\omega_{mn} \equiv \omega_{mn}(t)$ between initial and final anyon configurations.

To simulate the dynamics, we perform a stochastic unraveling of Eq.~\eqref{eq:rateEquation} following a standard time-dependent kinetic Monte Carlo approach. Specifically, we compute the time evolution using a ``first reaction'' algorithm~\cite{Gillespie1977} suitably modified for time-dependent rates~\cite{Jansen1995}. The basic idea is to evolve the system step by step by: (i) randomly drawing, for every possible spin flip $i$, a transition time $t_i$ from an exponential probability distribution $\gamma_i(\tau) \exp(-\int_{t_0}^{t_i} d\tau \gamma_i(\tau))$ (where $\gamma_i(\tau)$ is the transition rate at time $\tau$ and $t_0$ is the current simulation time), and (ii) by performing the spin flip $j$ with the minimum transition time $t_j = \min_i\{ t_i \}$ and updating the simulation time to $t'_0 = t_0 + t_j$. This procedure results in a single Monte Carlo ``run'', and averaging over multiple runs is required to faithfully reproduce the time evolution described by Eq.~\eqref{eq:rateEquation}. For a large number $N$ of runs, the statistical error decreases as $1/\sqrt{N}$. The error bars presented in our plots correspond to $68\%$ confidence intervals (one standard error of the mean).

To read out the state of the quantum memory, one typically performs a classical decoding step consisting of: (i) measuring the position of all anyons that are present (the ``error syndrome''), and (ii) using a classical algorithm to compute a correction operator that annihilates all anyons in pairs without introducing ``loop operators'' which would modify the memory state [see discussion below Eq.~\eqref{eq:toricCodeHam}]. In our simulations, we use a standard decoding algorithm based on ``minimal-weight perfect matching''~\cite{Edmonds1965,Dennis2002}, which has an efficient implementation called ``Blossom V''~\cite{Kolmogorov2009}.

We remark that minimal-weight perfect matching leads to an ``even-odd'' effect which must be accounted for to reproduce the numerical results obtained in Sec.~\ref{sec:simulationsCoherenceTime}. This is best understood from the point of view of a single anyon pair: In a system of odd size $L$, a logical error occurs when the anyons separate by a distance $(L+1)/2$. When $L$ is even, however, a logical error does not necessarily occur when anyons separate by a distance $L/2$. Since $L-(L/2) = L/2$, the decoder only fails half of the time in that case. Therefore, the probability that a logical error occurs is $[P_\mathrm{sep}(L/2) + P_\mathrm{sep}(L/2+1)]/2$, where $P_\mathrm{sep}(\ell)$ is the probability that the anyons separate by $\ell$.


\bibliographystyle{apsrev4-1}
\bibliography{manuscript}

\begin{thebibliography}{68}%
\makeatletter
\providecommand \@ifxundefined [1]{%
 \@ifx{#1\undefined}
}%
\providecommand \@ifnum [1]{%
 \ifnum #1\expandafter \@firstoftwo
 \else \expandafter \@secondoftwo
 \fi
}%
\providecommand \@ifx [1]{%
 \ifx #1\expandafter \@firstoftwo
 \else \expandafter \@secondoftwo
 \fi
}%
\providecommand \natexlab [1]{#1}%
\providecommand \enquote  [1]{``#1''}%
\providecommand \bibnamefont  [1]{#1}%
\providecommand \bibfnamefont [1]{#1}%
\providecommand \citenamefont [1]{#1}%
\providecommand \href@noop [0]{\@secondoftwo}%
\providecommand \href [0]{\begingroup \@sanitize@url \@href}%
\providecommand \@href[1]{\@@startlink{#1}\@@href}%
\providecommand \@@href[1]{\endgroup#1\@@endlink}%
\providecommand \@sanitize@url [0]{\catcode `\\12\catcode `\$12\catcode
  `\&12\catcode `\#12\catcode `\^12\catcode `\_12\catcode `\%12\relax}%
\providecommand \@@startlink[1]{}%
\providecommand \@@endlink[0]{}%
\providecommand \url  [0]{\begingroup\@sanitize@url \@url }%
\providecommand \@url [1]{\endgroup\@href {#1}{\urlprefix }}%
\providecommand \urlprefix  [0]{URL }%
\providecommand \Eprint [0]{\href }%
\providecommand \doibase [0]{http://dx.doi.org/}%
\providecommand \selectlanguage [0]{\@gobble}%
\providecommand \bibinfo  [0]{\@secondoftwo}%
\providecommand \bibfield  [0]{\@secondoftwo}%
\providecommand \translation [1]{[#1]}%
\providecommand \BibitemOpen [0]{}%
\providecommand \bibitemStop [0]{}%
\providecommand \bibitemNoStop [0]{.\EOS\space}%
\providecommand \EOS [0]{\spacefactor3000\relax}%
\providecommand \BibitemShut  [1]{\csname bibitem#1\endcsname}%
\let\auto@bib@innerbib\@empty
\bibitem [{\citenamefont {Dennis}\ \emph {et~al.}(2002)\citenamefont {Dennis},
  \citenamefont {Kitaev}, \citenamefont {Landahl},\ and\ \citenamefont
  {Preskill}}]{Dennis2002}%
  \BibitemOpen
  \bibfield  {author} {\bibinfo {author} {\bibfnamefont {E.}~\bibnamefont
  {Dennis}}, \bibinfo {author} {\bibfnamefont {A.}~\bibnamefont {Kitaev}},
  \bibinfo {author} {\bibfnamefont {A.}~\bibnamefont {Landahl}}, \ and\
  \bibinfo {author} {\bibfnamefont {J.}~\bibnamefont {Preskill}},\ }\bibfield
  {title} {\emph {\enquote {\bibinfo {title} {Topological quantum memory},}\
  }}\href {\doibase http://dx.doi.org/10.1063/1.1499754} {\bibfield  {journal}
  {\bibinfo  {journal} {Journal of Mathematical Physics}\ }\textbf {\bibinfo
  {volume} {43}},\ \bibinfo {pages} {4452} (\bibinfo {year}
  {2002})}\BibitemShut {NoStop}%
\bibitem [{\citenamefont {Bacon}(2006)}]{Bacon2006}%
  \BibitemOpen
  \bibfield  {author} {\bibinfo {author} {\bibfnamefont {D.}~\bibnamefont
  {Bacon}},\ }\bibfield  {title} {\emph {\enquote {\bibinfo {title} {Operator
  quantum error-correcting subsystems for self-correcting quantum memories},}\
  }}\href {\doibase 10.1103/PhysRevA.73.012340} {\bibfield  {journal} {\bibinfo
   {journal} {Phys. Rev. A}\ }\textbf {\bibinfo {volume} {73}},\ \bibinfo
  {pages} {012340} (\bibinfo {year} {2006})}\BibitemShut {NoStop}%
\bibitem [{\citenamefont {Bravyi}\ and\ \citenamefont
  {Terhal}(2009)}]{Bravyi2009}%
  \BibitemOpen
  \bibfield  {author} {\bibinfo {author} {\bibfnamefont {S.}~\bibnamefont
  {Bravyi}}\ and\ \bibinfo {author} {\bibfnamefont {B.}~\bibnamefont
  {Terhal}},\ }\bibfield  {title} {\emph {\enquote {\bibinfo {title} {A no-go
  theorem for a two-dimensional self-correcting quantum memory based on
  stabilizer codes},}\ }}\href
  {http://stacks.iop.org/1367-2630/11/i=4/a=043029} {\bibfield  {journal}
  {\bibinfo  {journal} {New Journal of Physics}\ }\textbf {\bibinfo {volume}
  {11}},\ \bibinfo {pages} {043029} (\bibinfo {year} {2009})}\BibitemShut
  {NoStop}%
\bibitem [{\citenamefont {Yoshida}(2011)}]{Yoshida2011}%
  \BibitemOpen
  \bibfield  {author} {\bibinfo {author} {\bibfnamefont {B.}~\bibnamefont
  {Yoshida}},\ }\bibfield  {title} {\emph {\enquote {\bibinfo {title}
  {Feasibility of self-correcting quantum memory and thermal stability of
  topological order},}\ }}\href {\doibase
  http://dx.doi.org/10.1016/j.aop.2011.06.001} {\bibfield  {journal} {\bibinfo
  {journal} {Annals of Physics}\ }\textbf {\bibinfo {volume} {326}},\ \bibinfo
  {pages} {2566} (\bibinfo {year} {2011})}\BibitemShut {NoStop}%
\bibitem [{\citenamefont {Alicki}\ \emph {et~al.}(2010)\citenamefont {Alicki},
  \citenamefont {Horodecki}, \citenamefont {Horodecki},\ and\ \citenamefont
  {Horodecki}}]{Alicki2010}%
  \BibitemOpen
  \bibfield  {author} {\bibinfo {author} {\bibfnamefont {R.}~\bibnamefont
  {Alicki}}, \bibinfo {author} {\bibfnamefont {M.}~\bibnamefont {Horodecki}},
  \bibinfo {author} {\bibfnamefont {P.}~\bibnamefont {Horodecki}}, \ and\
  \bibinfo {author} {\bibfnamefont {R.}~\bibnamefont {Horodecki}},\ }\bibfield
  {title} {\emph {\enquote {\bibinfo {title} {On thermal stability of
  topological qubit in kitaev's 4d model},}\ }}\href {\doibase
  10.1142/S1230161210000023} {\bibfield  {journal} {\bibinfo  {journal} {Open
  Syst. Inf. Dyn.}\ }\textbf {\bibinfo {volume} {17}},\ \bibinfo {pages} {1}
  (\bibinfo {year} {2010})}\BibitemShut {NoStop}%
\bibitem [{\citenamefont {Wen}(1990)}]{Wen1990}%
  \BibitemOpen
  \bibfield  {author} {\bibinfo {author} {\bibfnamefont {X.~G.}\ \bibnamefont
  {Wen}},\ }\bibfield  {title} {\emph {\enquote {\bibinfo {title} {Topological
  orders in rigid states},}\ }}\href {\doibase 10.1142/S0217979290000139}
  {\bibfield  {journal} {\bibinfo  {journal} {International Journal of Modern
  Physics B}\ }\textbf {\bibinfo {volume} {4}},\ \bibinfo {pages} {239}
  (\bibinfo {year} {1990})}\BibitemShut {NoStop}%
\bibitem [{\citenamefont {Bravyi}\ \emph {et~al.}(2010)\citenamefont {Bravyi},
  \citenamefont {Hastings},\ and\ \citenamefont {Michalakis}}]{Bravyi2010}%
  \BibitemOpen
  \bibfield  {author} {\bibinfo {author} {\bibfnamefont {S.}~\bibnamefont
  {Bravyi}}, \bibinfo {author} {\bibfnamefont {M.~B.}\ \bibnamefont
  {Hastings}}, \ and\ \bibinfo {author} {\bibfnamefont {S.}~\bibnamefont
  {Michalakis}},\ }\bibfield  {title} {\emph {\enquote {\bibinfo {title}
  {Topological quantum order: Stability under local perturbations},}\ }}\href
  {http://scitation.aip.org/content/aip/journal/jmp/51/9/10.1063/1.3490195}
  {\bibfield  {journal} {\bibinfo  {journal} {Journal of Mathematical Physics}\
  }\textbf {\bibinfo {volume} {51}} (\bibinfo {year} {2010})}\BibitemShut
  {NoStop}%
\bibitem [{\citenamefont {Michalakis}\ and\ \citenamefont
  {Zwolak}(2013)}]{Michalakis2013}%
  \BibitemOpen
  \bibfield  {author} {\bibinfo {author} {\bibfnamefont {S.}~\bibnamefont
  {Michalakis}}\ and\ \bibinfo {author} {\bibfnamefont {J.~P.}\ \bibnamefont
  {Zwolak}},\ }\bibfield  {title} {\emph {\enquote {\bibinfo {title} {Stability
  of frustration-free hamiltonians},}\ }}\href {\doibase
  10.1007/s00220-013-1762-6} {\bibfield  {journal} {\bibinfo  {journal}
  {Communications in Mathematical Physics}\ }\textbf {\bibinfo {volume}
  {322}},\ \bibinfo {pages} {277} (\bibinfo {year} {2013})}\BibitemShut
  {NoStop}%
\bibitem [{\citenamefont {Kitaev}(2003)}]{Kitaev2003}%
  \BibitemOpen
  \bibfield  {author} {\bibinfo {author} {\bibfnamefont {A.}~\bibnamefont
  {Kitaev}},\ }\bibfield  {title} {\emph {\enquote {\bibinfo {title}
  {Fault-tolerant quantum computation by anyons},}\ }}\href {\doibase
  http://dx.doi.org/10.1016/S0003-4916(02)00018-0} {\bibfield  {journal}
  {\bibinfo  {journal} {Annals of Physics}\ }\textbf {\bibinfo {volume}
  {303}},\ \bibinfo {pages} {2} (\bibinfo {year} {2003})}\BibitemShut {NoStop}%
\bibitem [{\citenamefont {Alicki}\ \emph {et~al.}(2009)\citenamefont {Alicki},
  \citenamefont {Fannes},\ and\ \citenamefont {Horodecki}}]{Alicki2009}%
  \BibitemOpen
  \bibfield  {author} {\bibinfo {author} {\bibfnamefont {R.}~\bibnamefont
  {Alicki}}, \bibinfo {author} {\bibfnamefont {M.}~\bibnamefont {Fannes}}, \
  and\ \bibinfo {author} {\bibfnamefont {M.}~\bibnamefont {Horodecki}},\
  }\bibfield  {title} {\emph {\enquote {\bibinfo {title} {On thermalization in
  kitaev's 2d model},}\ }}\href
  {http://stacks.iop.org/1751-8121/42/i=6/a=065303} {\bibfield  {journal}
  {\bibinfo  {journal} {Journal of Physics A: Mathematical and Theoretical}\
  }\textbf {\bibinfo {volume} {42}},\ \bibinfo {pages} {065303} (\bibinfo
  {year} {2009})}\BibitemShut {NoStop}%
\bibitem [{\citenamefont {Kay}\ and\ \citenamefont {Colbeck}(2008)}]{Kay2008}%
  \BibitemOpen
  \bibfield  {author} {\bibinfo {author} {\bibfnamefont {A.}~\bibnamefont
  {Kay}}\ and\ \bibinfo {author} {\bibfnamefont {R.}~\bibnamefont {Colbeck}},\
  }\bibfield  {title} {\emph {\enquote {\bibinfo {title} {Quantum
  self-correcting stabilizer codes},}\ }}\href {http://arxiv.org/abs/0810.3557}
  {\bibfield  {journal} {\bibinfo  {journal} {arXiv:0810.3557}\ } (\bibinfo
  {year} {2008})}\BibitemShut {NoStop}%
\bibitem [{\citenamefont {Pastawski}\ \emph {et~al.}(2010)\citenamefont
  {Pastawski}, \citenamefont {Kay}, \citenamefont {Schuch},\ and\ \citenamefont
  {Cirac}}]{Pastawski2010}%
  \BibitemOpen
  \bibfield  {author} {\bibinfo {author} {\bibfnamefont {F.}~\bibnamefont
  {Pastawski}}, \bibinfo {author} {\bibfnamefont {A.}~\bibnamefont {Kay}},
  \bibinfo {author} {\bibfnamefont {N.}~\bibnamefont {Schuch}}, \ and\ \bibinfo
  {author} {\bibfnamefont {I.}~\bibnamefont {Cirac}},\ }\bibfield  {title}
  {\emph {\enquote {\bibinfo {title} {Limitations of passive protection of
  quantum information},}\ }}\href
  {http://www.rintonpress.com/xxqic10/qic-10-78/0580-0618.pdf} {\bibfield
  {journal} {\bibinfo  {journal} {Quantum Information and Computation}\
  }\textbf {\bibinfo {volume} {10}},\ \bibinfo {pages} {0580} (\bibinfo {year}
  {2010})}\BibitemShut {NoStop}%
\bibitem [{\citenamefont {Landon-Cardinal}\ and\ \citenamefont
  {Poulin}(2013)}]{Landon-Cardinal2013}%
  \BibitemOpen
  \bibfield  {author} {\bibinfo {author} {\bibfnamefont {O.}~\bibnamefont
  {Landon-Cardinal}}\ and\ \bibinfo {author} {\bibfnamefont {D.}~\bibnamefont
  {Poulin}},\ }\bibfield  {title} {\emph {\enquote {\bibinfo {title} {Local
  topological order inhibits thermal stability in 2d},}\ }}\href {\doibase
  10.1103/PhysRevLett.110.090502} {\bibfield  {journal} {\bibinfo  {journal}
  {Phys. Rev. Lett.}\ }\textbf {\bibinfo {volume} {110}},\ \bibinfo {pages}
  {090502} (\bibinfo {year} {2013})}\BibitemShut {NoStop}%
\bibitem [{\citenamefont {Pastawski}\ and\ \citenamefont
  {Yoshida}(2015)}]{Pastawski2015}%
  \BibitemOpen
  \bibfield  {author} {\bibinfo {author} {\bibfnamefont {F.}~\bibnamefont
  {Pastawski}}\ and\ \bibinfo {author} {\bibfnamefont {B.}~\bibnamefont
  {Yoshida}},\ }\bibfield  {title} {\emph {\enquote {\bibinfo {title}
  {Fault-tolerant logical gates in quantum error-correcting codes},}\ }}\href
  {\doibase 10.1103/PhysRevA.91.012305} {\bibfield  {journal} {\bibinfo
  {journal} {Phys. Rev. A}\ }\textbf {\bibinfo {volume} {91}},\ \bibinfo
  {pages} {012305} (\bibinfo {year} {2015})}\BibitemShut {NoStop}%
\bibitem [{\citenamefont {Hamma}\ \emph {et~al.}(2009)\citenamefont {Hamma},
  \citenamefont {Castelnovo},\ and\ \citenamefont {Chamon}}]{Hamma2009}%
  \BibitemOpen
  \bibfield  {author} {\bibinfo {author} {\bibfnamefont {A.}~\bibnamefont
  {Hamma}}, \bibinfo {author} {\bibfnamefont {C.}~\bibnamefont {Castelnovo}}, \
  and\ \bibinfo {author} {\bibfnamefont {C.}~\bibnamefont {Chamon}},\
  }\bibfield  {title} {\emph {\enquote {\bibinfo {title} {Toric-boson model:
  Toward a topological quantum memory at finite temperature},}\ }}\href
  {\doibase 10.1103/PhysRevB.79.245122} {\bibfield  {journal} {\bibinfo
  {journal} {Phys. Rev. B}\ }\textbf {\bibinfo {volume} {79}},\ \bibinfo
  {pages} {245122} (\bibinfo {year} {2009})}\BibitemShut {NoStop}%
\bibitem [{\citenamefont {Chesi}\ \emph {et~al.}(2010)\citenamefont {Chesi},
  \citenamefont {R\"othlisberger},\ and\ \citenamefont {Loss}}]{Chesi2010}%
  \BibitemOpen
  \bibfield  {author} {\bibinfo {author} {\bibfnamefont {S.}~\bibnamefont
  {Chesi}}, \bibinfo {author} {\bibfnamefont {B.}~\bibnamefont
  {R\"othlisberger}}, \ and\ \bibinfo {author} {\bibfnamefont {D.}~\bibnamefont
  {Loss}},\ }\bibfield  {title} {\emph {\enquote {\bibinfo {title}
  {Self-correcting quantum memory in a thermal environment},}\ }}\href
  {\doibase 10.1103/PhysRevA.82.022305} {\bibfield  {journal} {\bibinfo
  {journal} {Phys. Rev. A}\ }\textbf {\bibinfo {volume} {82}},\ \bibinfo
  {pages} {022305} (\bibinfo {year} {2010})}\BibitemShut {NoStop}%
\bibitem [{\citenamefont {Pedrocchi}\ \emph {et~al.}(2013)\citenamefont
  {Pedrocchi}, \citenamefont {Hutter}, \citenamefont {Wootton},\ and\
  \citenamefont {Loss}}]{Pedrocchi2013}%
  \BibitemOpen
  \bibfield  {author} {\bibinfo {author} {\bibfnamefont {F.~L.}\ \bibnamefont
  {Pedrocchi}}, \bibinfo {author} {\bibfnamefont {A.}~\bibnamefont {Hutter}},
  \bibinfo {author} {\bibfnamefont {J.~R.}\ \bibnamefont {Wootton}}, \ and\
  \bibinfo {author} {\bibfnamefont {D.}~\bibnamefont {Loss}},\ }\bibfield
  {title} {\emph {\enquote {\bibinfo {title} {Enhanced thermal stability of the
  toric code through coupling to a bosonic bath},}\ }}\href {\doibase
  10.1103/PhysRevA.88.062313} {\bibfield  {journal} {\bibinfo  {journal} {Phys.
  Rev. A}\ }\textbf {\bibinfo {volume} {88}},\ \bibinfo {pages} {062313}
  (\bibinfo {year} {2013})}\BibitemShut {NoStop}%
\bibitem [{\citenamefont {Hutter}\ \emph {et~al.}(2014)\citenamefont {Hutter},
  \citenamefont {Pedrocchi}, \citenamefont {Wootton},\ and\ \citenamefont
  {Loss}}]{Hutter2014}%
  \BibitemOpen
  \bibfield  {author} {\bibinfo {author} {\bibfnamefont {A.}~\bibnamefont
  {Hutter}}, \bibinfo {author} {\bibfnamefont {F.~L.}\ \bibnamefont
  {Pedrocchi}}, \bibinfo {author} {\bibfnamefont {J.~R.}\ \bibnamefont
  {Wootton}}, \ and\ \bibinfo {author} {\bibfnamefont {D.}~\bibnamefont
  {Loss}},\ }\bibfield  {title} {\emph {\enquote {\bibinfo {title} {Effective
  quantum-memory hamiltonian from local two-body interactions},}\ }}\href
  {\doibase 10.1103/PhysRevA.90.012321} {\bibfield  {journal} {\bibinfo
  {journal} {Phys. Rev. A}\ }\textbf {\bibinfo {volume} {90}},\ \bibinfo
  {pages} {012321} (\bibinfo {year} {2014})}\BibitemShut {NoStop}%
\bibitem [{\citenamefont {Haah}(2011)}]{Haah2011}%
  \BibitemOpen
  \bibfield  {author} {\bibinfo {author} {\bibfnamefont {J.}~\bibnamefont
  {Haah}},\ }\bibfield  {title} {\emph {\enquote {\bibinfo {title} {Local
  stabilizer codes in three dimensions without string logical operators},}\
  }}\href {\doibase 10.1103/PhysRevA.83.042330} {\bibfield  {journal} {\bibinfo
   {journal} {Phys. Rev. A}\ }\textbf {\bibinfo {volume} {83}},\ \bibinfo
  {pages} {042330} (\bibinfo {year} {2011})}\BibitemShut {NoStop}%
\bibitem [{\citenamefont {Bravyi}\ and\ \citenamefont
  {Haah}(2011)}]{Bravyi2011}%
  \BibitemOpen
  \bibfield  {author} {\bibinfo {author} {\bibfnamefont {S.}~\bibnamefont
  {Bravyi}}\ and\ \bibinfo {author} {\bibfnamefont {J.}~\bibnamefont {Haah}},\
  }\bibfield  {title} {\emph {\enquote {\bibinfo {title} {Energy landscape of
  3d spin hamiltonians with topological order},}\ }}\href {\doibase
  10.1103/PhysRevLett.107.150504} {\bibfield  {journal} {\bibinfo  {journal}
  {Phys. Rev. Lett.}\ }\textbf {\bibinfo {volume} {107}},\ \bibinfo {pages}
  {150504} (\bibinfo {year} {2011})}\BibitemShut {NoStop}%
\bibitem [{\citenamefont {Bravyi}\ and\ \citenamefont
  {Haah}(2013)}]{Bravyi2013}%
  \BibitemOpen
  \bibfield  {author} {\bibinfo {author} {\bibfnamefont {S.}~\bibnamefont
  {Bravyi}}\ and\ \bibinfo {author} {\bibfnamefont {J.}~\bibnamefont {Haah}},\
  }\bibfield  {title} {\emph {\enquote {\bibinfo {title} {Quantum
  self-correction in the 3d cubic code model},}\ }}\href {\doibase
  10.1103/PhysRevLett.111.200501} {\bibfield  {journal} {\bibinfo  {journal}
  {Phys. Rev. Lett.}\ }\textbf {\bibinfo {volume} {111}},\ \bibinfo {pages}
  {200501} (\bibinfo {year} {2013})}\BibitemShut {NoStop}%
\bibitem [{\citenamefont {Kim}(2012)}]{Kim2012}%
  \BibitemOpen
  \bibfield  {author} {\bibinfo {author} {\bibfnamefont {I.~H.}\ \bibnamefont
  {Kim}},\ }\bibfield  {title} {\emph {\enquote {\bibinfo {title} {3d local
  qupit quantum code without string logical operator},}\ }}\href
  {http://arxiv.org/abs/1202.0052} {\bibfield  {journal} {\bibinfo  {journal}
  {arXiv:1202.0052}\ } (\bibinfo {year} {2012})}\BibitemShut {NoStop}%
\bibitem [{\citenamefont {Michnicki}(2012)}]{Michnicki2012}%
  \BibitemOpen
  \bibfield  {author} {\bibinfo {author} {\bibfnamefont {K.~P.}\ \bibnamefont
  {Michnicki}},\ }\bibfield  {title} {\emph {\enquote {\bibinfo {title} {3-d
  quantum stabilizer codes with a power law energy barrier},}\ }}\href
  {http://arxiv.org/abs/1208.3496} {\bibfield  {journal} {\bibinfo  {journal}
  {arXiv:1208.3496}\ } (\bibinfo {year} {2012})}\BibitemShut {NoStop}%
\bibitem [{\citenamefont {Michnicki}(2014)}]{Michnicki2014}%
  \BibitemOpen
  \bibfield  {author} {\bibinfo {author} {\bibfnamefont {K.~P.}\ \bibnamefont
  {Michnicki}},\ }\bibfield  {title} {\emph {\enquote {\bibinfo {title} {3d
  topological quantum memory with a power-law energy barrier},}\ }}\href
  {\doibase 10.1103/PhysRevLett.113.130501} {\bibfield  {journal} {\bibinfo
  {journal} {Phys. Rev. Lett.}\ }\textbf {\bibinfo {volume} {113}},\ \bibinfo
  {pages} {130501} (\bibinfo {year} {2014})}\BibitemShut {NoStop}%
\bibitem [{\citenamefont {Brown}\ \emph
  {et~al.}(2014{\natexlab{a}})\citenamefont {Brown}, \citenamefont
  {Al-Shimary},\ and\ \citenamefont {Pachos}}]{Brown2014}%
  \BibitemOpen
  \bibfield  {author} {\bibinfo {author} {\bibfnamefont {B.~J.}\ \bibnamefont
  {Brown}}, \bibinfo {author} {\bibfnamefont {A.}~\bibnamefont {Al-Shimary}}, \
  and\ \bibinfo {author} {\bibfnamefont {J.~K.}\ \bibnamefont {Pachos}},\
  }\bibfield  {title} {\emph {\enquote {\bibinfo {title} {Entropic barriers for
  two-dimensional quantum memories},}\ }}\href {\doibase
  10.1103/PhysRevLett.112.120503} {\bibfield  {journal} {\bibinfo  {journal}
  {Phys. Rev. Lett.}\ }\textbf {\bibinfo {volume} {112}},\ \bibinfo {pages}
  {120503} (\bibinfo {year} {2014}{\natexlab{a}})}\BibitemShut {NoStop}%
\bibitem [{\citenamefont {Stark}\ \emph {et~al.}(2011)\citenamefont {Stark},
  \citenamefont {Pollet}, \citenamefont {Imamo\ifmmode~\breve{g}\else
  \u{g}\fi{}lu},\ and\ \citenamefont {Renner}}]{Stark2011}%
  \BibitemOpen
  \bibfield  {author} {\bibinfo {author} {\bibfnamefont {C.}~\bibnamefont
  {Stark}}, \bibinfo {author} {\bibfnamefont {L.}~\bibnamefont {Pollet}},
  \bibinfo {author} {\bibfnamefont {A.~m.~c.}\ \bibnamefont
  {Imamo\ifmmode~\breve{g}\else \u{g}\fi{}lu}}, \ and\ \bibinfo {author}
  {\bibfnamefont {R.}~\bibnamefont {Renner}},\ }\bibfield  {title} {\emph
  {\enquote {\bibinfo {title} {Localization of toric code defects},}\ }}\href
  {\doibase 10.1103/PhysRevLett.107.030504} {\bibfield  {journal} {\bibinfo
  {journal} {Phys. Rev. Lett.}\ }\textbf {\bibinfo {volume} {107}},\ \bibinfo
  {pages} {030504} (\bibinfo {year} {2011})}\BibitemShut {NoStop}%
\bibitem [{\citenamefont {Wootton}\ and\ \citenamefont
  {Pachos}(2011)}]{Wootton2011}%
  \BibitemOpen
  \bibfield  {author} {\bibinfo {author} {\bibfnamefont {J.~R.}\ \bibnamefont
  {Wootton}}\ and\ \bibinfo {author} {\bibfnamefont {J.~K.}\ \bibnamefont
  {Pachos}},\ }\bibfield  {title} {\emph {\enquote {\bibinfo {title} {Bringing
  order through disorder: Localization of errors in topological quantum
  memories},}\ }}\href {\doibase 10.1103/PhysRevLett.107.030503} {\bibfield
  {journal} {\bibinfo  {journal} {Phys. Rev. Lett.}\ }\textbf {\bibinfo
  {volume} {107}},\ \bibinfo {pages} {030503} (\bibinfo {year}
  {2011})}\BibitemShut {NoStop}%
\bibitem [{\citenamefont {Bravyi}\ and\ \citenamefont
  {K\"onig}(2012)}]{Bravyi2012}%
  \BibitemOpen
  \bibfield  {author} {\bibinfo {author} {\bibfnamefont {S.}~\bibnamefont
  {Bravyi}}\ and\ \bibinfo {author} {\bibfnamefont {R.}~\bibnamefont
  {K\"onig}},\ }\bibfield  {title} {\emph {\enquote {\bibinfo {title}
  {Disorder-assisted error correction in majorana chains},}\ }}\href {\doibase
  10.1007/s00220-012-1606-9} {\bibfield  {journal} {\bibinfo  {journal}
  {Communications in Mathematical Physics}\ }\textbf {\bibinfo {volume}
  {316}},\ \bibinfo {pages} {641} (\bibinfo {year} {2012})}\BibitemShut
  {NoStop}%
\bibitem [{\citenamefont {Pastawski}\ \emph {et~al.}(2011)\citenamefont
  {Pastawski}, \citenamefont {Clemente},\ and\ \citenamefont
  {Cirac}}]{Pastawski2011}%
  \BibitemOpen
  \bibfield  {author} {\bibinfo {author} {\bibfnamefont {F.}~\bibnamefont
  {Pastawski}}, \bibinfo {author} {\bibfnamefont {L.}~\bibnamefont {Clemente}},
  \ and\ \bibinfo {author} {\bibfnamefont {J.~I.}\ \bibnamefont {Cirac}},\
  }\bibfield  {title} {\emph {\enquote {\bibinfo {title} {Quantum memories
  based on engineered dissipation},}\ }}\href {\doibase
  10.1103/PhysRevA.83.012304} {\bibfield  {journal} {\bibinfo  {journal} {Phys.
  Rev. A}\ }\textbf {\bibinfo {volume} {83}},\ \bibinfo {pages} {012304}
  (\bibinfo {year} {2011})}\BibitemShut {NoStop}%
\bibitem [{\citenamefont {Fujii}\ \emph {et~al.}(2014)\citenamefont {Fujii},
  \citenamefont {Negoro}, \citenamefont {Imoto},\ and\ \citenamefont
  {Kitagawa}}]{Fujii2014}%
  \BibitemOpen
  \bibfield  {author} {\bibinfo {author} {\bibfnamefont {K.}~\bibnamefont
  {Fujii}}, \bibinfo {author} {\bibfnamefont {M.}~\bibnamefont {Negoro}},
  \bibinfo {author} {\bibfnamefont {N.}~\bibnamefont {Imoto}}, \ and\ \bibinfo
  {author} {\bibfnamefont {M.}~\bibnamefont {Kitagawa}},\ }\bibfield  {title}
  {\emph {\enquote {\bibinfo {title} {Measurement-free topological protection
  using dissipative feedback},}\ }}\href {\doibase 10.1103/PhysRevX.4.041039}
  {\bibfield  {journal} {\bibinfo  {journal} {Phys. Rev. X}\ }\textbf {\bibinfo
  {volume} {4}},\ \bibinfo {pages} {041039} (\bibinfo {year}
  {2014})}\BibitemShut {NoStop}%
\bibitem [{\citenamefont {Kapit}\ \emph {et~al.}(2015)\citenamefont {Kapit},
  \citenamefont {Chalker},\ and\ \citenamefont {Simon}}]{Kapit2015}%
  \BibitemOpen
  \bibfield  {author} {\bibinfo {author} {\bibfnamefont {E.}~\bibnamefont
  {Kapit}}, \bibinfo {author} {\bibfnamefont {J.~T.}\ \bibnamefont {Chalker}},
  \ and\ \bibinfo {author} {\bibfnamefont {S.~H.}\ \bibnamefont {Simon}},\
  }\bibfield  {title} {\emph {\enquote {\bibinfo {title} {Passive correction of
  quantum logical errors in a driven, dissipative system: A blueprint for an
  analog quantum code fabric},}\ }}\href {\doibase 10.1103/PhysRevA.91.062324}
  {\bibfield  {journal} {\bibinfo  {journal} {Phys. Rev. A}\ }\textbf {\bibinfo
  {volume} {91}},\ \bibinfo {pages} {062324} (\bibinfo {year}
  {2015})}\BibitemShut {NoStop}%
\bibitem [{\citenamefont {Brown}\ \emph
  {et~al.}(2014{\natexlab{b}})\citenamefont {Brown}, \citenamefont {Loss},
  \citenamefont {Pachos}, \citenamefont {Self},\ and\ \citenamefont
  {Wootton}}]{Brown2014_2}%
  \BibitemOpen
  \bibfield  {author} {\bibinfo {author} {\bibfnamefont {B.~J.}\ \bibnamefont
  {Brown}}, \bibinfo {author} {\bibfnamefont {D.}~\bibnamefont {Loss}},
  \bibinfo {author} {\bibfnamefont {J.~K.}\ \bibnamefont {Pachos}}, \bibinfo
  {author} {\bibfnamefont {C.~N.}\ \bibnamefont {Self}}, \ and\ \bibinfo
  {author} {\bibfnamefont {J.~R.}\ \bibnamefont {Wootton}},\ }\bibfield
  {title} {\emph {\enquote {\bibinfo {title} {Quantum memories at finite
  temperature},}\ }}\href {http://arxiv.org/abs/1411.6643} {\bibfield
  {journal} {\bibinfo  {journal} {arXiv:1411.6643}\ } (\bibinfo {year}
  {2014}{\natexlab{b}})}\BibitemShut {NoStop}%
\bibitem [{\citenamefont {Landon-Cardinal}\ \emph {et~al.}(2015)\citenamefont
  {Landon-Cardinal}, \citenamefont {Yoshida}, \citenamefont {Poulin},\ and\
  \citenamefont {Preskill}}]{Landon-Cardinal2015}%
  \BibitemOpen
  \bibfield  {author} {\bibinfo {author} {\bibfnamefont {O.}~\bibnamefont
  {Landon-Cardinal}}, \bibinfo {author} {\bibfnamefont {B.}~\bibnamefont
  {Yoshida}}, \bibinfo {author} {\bibfnamefont {D.}~\bibnamefont {Poulin}}, \
  and\ \bibinfo {author} {\bibfnamefont {J.}~\bibnamefont {Preskill}},\
  }\bibfield  {title} {\emph {\enquote {\bibinfo {title} {Perturbative
  instability of quantum memory based on effective long-range interactions},}\
  }}\href {\doibase 10.1103/PhysRevA.91.032303} {\bibfield  {journal} {\bibinfo
   {journal} {Phys. Rev. A}\ }\textbf {\bibinfo {volume} {91}},\ \bibinfo
  {pages} {032303} (\bibinfo {year} {2015})}\BibitemShut {NoStop}%
\bibitem [{\citenamefont {Terhal}(2015)}]{Terhal2015}%
  \BibitemOpen
  \bibfield  {author} {\bibinfo {author} {\bibfnamefont {B.~M.}\ \bibnamefont
  {Terhal}},\ }\bibfield  {title} {\emph {\enquote {\bibinfo {title} {Quantum
  error correction for quantum memories},}\ }}\href {\doibase
  10.1103/RevModPhys.87.307} {\bibfield  {journal} {\bibinfo  {journal} {Rev.
  Mod. Phys.}\ }\textbf {\bibinfo {volume} {87}},\ \bibinfo {pages} {307}
  (\bibinfo {year} {2015})}\BibitemShut {NoStop}%
\bibitem [{\citenamefont {Herold}\ \emph
  {et~al.}(2015{\natexlab{a}})\citenamefont {Herold}, \citenamefont {Campbell},
  \citenamefont {Eisert},\ and\ \citenamefont {Kastoryano}}]{Herold2015}%
  \BibitemOpen
  \bibfield  {author} {\bibinfo {author} {\bibfnamefont {M.}~\bibnamefont
  {Herold}}, \bibinfo {author} {\bibfnamefont {E.~T.}\ \bibnamefont
  {Campbell}}, \bibinfo {author} {\bibfnamefont {J.}~\bibnamefont {Eisert}}, \
  and\ \bibinfo {author} {\bibfnamefont {M.~J.}\ \bibnamefont {Kastoryano}},\
  }\bibfield  {title} {\emph {\enquote {\bibinfo {title} {Cellular-automaton
  decoders for topological quantum memories},}\ }}\href {\doibase
  10.1038/npjqi.2015.10} {\bibfield  {journal} {\bibinfo  {journal} {npj Quant.
  Inf.}\ }\textbf {\bibinfo {volume} {1}},\ \bibinfo {pages} {15010} (\bibinfo
  {year} {2015}{\natexlab{a}})}\BibitemShut {NoStop}%
\bibitem [{\citenamefont {Herold}\ \emph
  {et~al.}(2015{\natexlab{b}})\citenamefont {Herold}, \citenamefont
  {Kastoryano}, \citenamefont {Campbell},\ and\ \citenamefont
  {Eisert}}]{Herold2015_2}%
  \BibitemOpen
  \bibfield  {author} {\bibinfo {author} {\bibfnamefont {M.}~\bibnamefont
  {Herold}}, \bibinfo {author} {\bibfnamefont {M.~J.}\ \bibnamefont
  {Kastoryano}}, \bibinfo {author} {\bibfnamefont {E.~T.}\ \bibnamefont
  {Campbell}}, \ and\ \bibinfo {author} {\bibfnamefont {J.}~\bibnamefont
  {Eisert}},\ }\bibfield  {title} {\emph {\enquote {\bibinfo {title} {Fault
  tolerant dynamical decoders for topological quantum memories},}\ }}\href
  {http://arxiv.org/abs/1511.05579} {\bibfield  {journal} {\bibinfo  {journal}
  {arXiv:1511.05579}\ } (\bibinfo {year} {2015}{\natexlab{b}})}\BibitemShut
  {NoStop}%
\bibitem [{\citenamefont {Gottesman}(1997)}]{Gottesman1997}%
  \BibitemOpen
  \bibfield  {author} {\bibinfo {author} {\bibfnamefont {D.}~\bibnamefont
  {Gottesman}},\ }\bibfield  {title} {\emph {\enquote {\bibinfo {title}
  {Stabilizer codes and quantum error correction},}\ }}\href
  {http://arxiv.org/abs/quant-ph/9705052} {\bibfield  {journal} {\bibinfo
  {journal} {arXiv:quant-ph/9705052}\ } (\bibinfo {year} {1997})}\BibitemShut
  {NoStop}%
\bibitem [{\citenamefont {Leggett}\ \emph {et~al.}(1987)\citenamefont
  {Leggett}, \citenamefont {Chakravarty}, \citenamefont {Dorsey}, \citenamefont
  {Fisher}, \citenamefont {Garg},\ and\ \citenamefont {Zwerger}}]{Leggett1987}%
  \BibitemOpen
  \bibfield  {author} {\bibinfo {author} {\bibfnamefont {A.~J.}\ \bibnamefont
  {Leggett}}, \bibinfo {author} {\bibfnamefont {S.}~\bibnamefont
  {Chakravarty}}, \bibinfo {author} {\bibfnamefont {A.~T.}\ \bibnamefont
  {Dorsey}}, \bibinfo {author} {\bibfnamefont {M.~P.~A.}\ \bibnamefont
  {Fisher}}, \bibinfo {author} {\bibfnamefont {A.}~\bibnamefont {Garg}}, \ and\
  \bibinfo {author} {\bibfnamefont {W.}~\bibnamefont {Zwerger}},\ }\bibfield
  {title} {\emph {\enquote {\bibinfo {title} {Dynamics of the dissipative
  two-state system},}\ }}\href {\doibase 10.1103/RevModPhys.59.1} {\bibfield
  {journal} {\bibinfo  {journal} {Rev. Mod. Phys.}\ }\textbf {\bibinfo {volume}
  {59}},\ \bibinfo {pages} {1} (\bibinfo {year} {1987})}\BibitemShut {NoStop}%
\bibitem [{\citenamefont {Castelnovo}\ and\ \citenamefont
  {Chamon}(2011)}]{Castelnovo2011}%
  \BibitemOpen
  \bibfield  {author} {\bibinfo {author} {\bibfnamefont {C.}~\bibnamefont
  {Castelnovo}}\ and\ \bibinfo {author} {\bibfnamefont {C.}~\bibnamefont
  {Chamon}},\ }\bibfield  {title} {\emph {\enquote {\bibinfo {title}
  {Topological quantum glassiness},}\ }}\href {\doibase
  10.1080/14786435.2011.609152} {\bibfield  {journal} {\bibinfo  {journal}
  {Phil. Mag.}\ }\textbf {\bibinfo {volume} {92}},\ \bibinfo {pages} {304}
  (\bibinfo {year} {2011})}\BibitemShut {NoStop}%
\bibitem [{\citenamefont {Haah}(2013)}]{Haah2013}%
  \BibitemOpen
  \bibfield  {author} {\bibinfo {author} {\bibfnamefont {J.}~\bibnamefont
  {Haah}},\ }\bibfield  {title} {\emph {\enquote {\bibinfo {title} {Commuting
  pauli hamiltonians as maps between free modules},}\ }}\href {\doibase
  10.1007/s00220-013-1810-2} {\bibfield  {journal} {\bibinfo  {journal}
  {Commun. Math. Phys.}\ }\textbf {\bibinfo {volume} {324}},\ \bibinfo {pages}
  {351} (\bibinfo {year} {2013})}\BibitemShut {NoStop}%
\bibitem [{\citenamefont {Brell}(2014)}]{Brell2014}%
  \BibitemOpen
  \bibfield  {author} {\bibinfo {author} {\bibfnamefont {C.~G.}\ \bibnamefont
  {Brell}},\ }\bibfield  {title} {\emph {\enquote {\bibinfo {title} {A proposal
  for self-correcting stabilizer quantum memories in 3 dimensions (or slightly
  less)},}\ }}\href {http://arxiv.org/abs/1411.7046} {\bibfield  {journal}
  {\bibinfo  {journal} {arXiv:1411.7046}\ } (\bibinfo {year}
  {2014})}\BibitemShut {NoStop}%
\bibitem [{Note1()}]{Note1}%
  \BibitemOpen
  \bibinfo {note} {As discussed in Sec.~\ref {sec:conclusion}, our scheme could
  also be used for active error correction or decoding.}\BibitemShut {Stop}%
\bibitem [{\citenamefont {Breuer}\ and\ \citenamefont
  {Petruccione}(2007)}]{Breuer2007}%
  \BibitemOpen
  \bibfield  {author} {\bibinfo {author} {\bibfnamefont {H.~P.}\ \bibnamefont
  {Breuer}}\ and\ \bibinfo {author} {\bibfnamefont {F.}~\bibnamefont
  {Petruccione}},\ }\href@noop {} {\emph {\bibinfo {title} {The Theory of Open
  Quantum Systems}}}\ (\bibinfo  {publisher} {Oxford University Press, USA},\
  \bibinfo {year} {2007})\BibitemShut {NoStop}%
\bibitem [{Note2()}]{Note2}%
  \BibitemOpen
  \bibinfo {note} {In particular, the probability that an isolated anyon
  located at the end of a 1D trench ``jumps out'' of the latter (thereby
  extending it) is equal to $\gamma (U)/[3\gamma (U) + \gamma (0)] = 1/(3\alpha
  + 1)$.}\BibitemShut {Stop}%
\bibitem [{Note3()}]{Note3}%
  \BibitemOpen
  \bibinfo {note} {Assuming that the anyon sinks to its potential minimum $-U$
  after a (rare) escape event, the probability that it jumps even further from
  its original trench reads $\DOTSI \intop \ilimits@ _0^{1/\gamma _\protect
  \mathrm {pump}} \protect \mathrm {d}\tau (\gamma [U(\tau )]/P_\protect
  \mathrm {esc})[\gamma (U)/(\gamma (U) + \gamma [U-U(\tau )])]$, which yields
  $P_2 = 1/4$ in the limit $U \gg T$}\BibitemShut {NoStop}%
\bibitem [{Note4()}]{Note4}%
  \BibitemOpen
  \bibinfo {note} {The rate at which a potential trench of size $\ell $ is
  extended is $\sim \gamma (U)/\ell $, where $1/\ell $ is the approximate
  probability that an anyon is located at the trench-potential wall. The total
  time required to extend a trench from initial size $2$ to $L/2$ is thus $\tau
  _\protect \mathrm {sep} \sim \DOTSB \sum@ \slimits@ _{\ell = 2}^{L/2} \gamma
  (U)^{-1} \ell \sim \gamma (U)^{-1} L^2$, yielding $\tau _\protect \mathrm
  {sep} \sim \gamma (U)/L^2$ (or $\sim \gamma (U)/L^3$ for 2D
  trenches).}\BibitemShut {Stop}%
\bibitem [{Note5()}]{Note5}%
  \BibitemOpen
  \bibinfo {note} {Anyon pairs can also be created partly inside an existing
  trench with rate $\sim \ell \gamma (2J-U)$. Such events can be neglected as
  compared to anyon-pair creations inside a trench, which occur at a faster
  rate [in the low-temperature regime of interest where $T \ll U$, such that
  $\gamma (2J-U) \ll \gamma (2J-2U)$].}\BibitemShut {Stop}%
\bibitem [{Note6()}]{Note6}%
  \BibitemOpen
  \bibinfo {note} {The probability that diffusing anyons separate by a distance
  $\ell $ without recombining is $\sim 1/\ell $ in 1D [see above Eq.~\protect
  \textup {\hbox {\mathsurround \z@ \protect \normalfont (\ignorespaces \ref
  {eq:1DResult}\unskip \@@italiccorr )}}], and $\sim 1/\protect \qopname \relax
  o{log}(\ell )$ in 2D (see, e.g., Ref.~\cite {Brown2014_2}). We neglect
  non-essential corrections coming from the bias towards annihilation when two
  anyons meet.}\BibitemShut {Stop}%
\bibitem [{Note7()}]{Note7}%
  \BibitemOpen
  \bibinfo {note} {As mentioned above Eq.~\protect \textup {\hbox
  {\mathsurround \z@ \protect \normalfont (\ignorespaces \ref {eq:rhot}\unskip
  \@@italiccorr )}}, $(L/2)^{\eta -\eta '}$ should be replaced by
  $(L/2)^2/\protect \qopname \relax o{log}(L/2)$ for 2D trenches.}\BibitemShut
  {Stop}%
\bibitem [{Note8()}]{Note8}%
  \BibitemOpen
  \bibinfo {note} {In the regime of interest where $(J-U)/T \gg 1$, $\tau
  _\protect \mathrm {coh}^\protect \mathrm {max}$ increases with $U/T$ despite
  the exponential factor $\protect \mathrm {e}^{\beta (J-U)/2}$ in the double
  exponential.}\BibitemShut {Stop}%
\bibitem [{\citenamefont {M\"uller}\ \emph {et~al.}(2011)\citenamefont
  {M\"uller}, \citenamefont {Hammerer}, \citenamefont {Zhou}, \citenamefont
  {Roos},\ and\ \citenamefont {Zoller}}]{Muller2011}%
  \BibitemOpen
  \bibfield  {author} {\bibinfo {author} {\bibfnamefont {M.}~\bibnamefont
  {M\"uller}}, \bibinfo {author} {\bibfnamefont {K.}~\bibnamefont {Hammerer}},
  \bibinfo {author} {\bibfnamefont {Y.~L.}\ \bibnamefont {Zhou}}, \bibinfo
  {author} {\bibfnamefont {C.~F.}\ \bibnamefont {Roos}}, \ and\ \bibinfo
  {author} {\bibfnamefont {P.}~\bibnamefont {Zoller}},\ }\bibfield  {title}
  {\emph {\enquote {\bibinfo {title} {Simulating open quantum systems: from
  many-body interactions to stabilizer pumping},}\ }}\href
  {http://stacks.iop.org/1367-2630/13/i=8/a=085007} {\bibfield  {journal}
  {\bibinfo  {journal} {New Journal of Physics}\ }\textbf {\bibinfo {volume}
  {13}},\ \bibinfo {pages} {085007} (\bibinfo {year} {2011})}\BibitemShut
  {NoStop}%
\bibitem [{\citenamefont {Fowler}\ \emph {et~al.}(2012)\citenamefont {Fowler},
  \citenamefont {Mariantoni}, \citenamefont {Martinis},\ and\ \citenamefont
  {Cleland}}]{Fowler2012}%
  \BibitemOpen
  \bibfield  {author} {\bibinfo {author} {\bibfnamefont {A.~G.}\ \bibnamefont
  {Fowler}}, \bibinfo {author} {\bibfnamefont {M.}~\bibnamefont {Mariantoni}},
  \bibinfo {author} {\bibfnamefont {J.~M.}\ \bibnamefont {Martinis}}, \ and\
  \bibinfo {author} {\bibfnamefont {A.~N.}\ \bibnamefont {Cleland}},\
  }\bibfield  {title} {\emph {\enquote {\bibinfo {title} {Surface codes:
  Towards practical large-scale quantum computation},}\ }}\href {\doibase
  10.1103/PhysRevA.86.032324} {\bibfield  {journal} {\bibinfo  {journal} {Phys.
  Rev. A}\ }\textbf {\bibinfo {volume} {86}},\ \bibinfo {pages} {032324}
  (\bibinfo {year} {2012})}\BibitemShut {NoStop}%
\bibitem [{\citenamefont {Kelly}\ \emph {et~al.}(2015)\citenamefont {Kelly},
  \citenamefont {Barends}, \citenamefont {Fowler}, \citenamefont {Megrant},
  \citenamefont {Jeffrey}, \citenamefont {White}, \citenamefont {Sank},
  \citenamefont {Mutus}, \citenamefont {Campbell}, \citenamefont {Chen},
  \citenamefont {Chen}, \citenamefont {Chiaro}, \citenamefont {Dunsworth},
  \citenamefont {Hoi}, \citenamefont {Neill}, \citenamefont {O/'Malley},
  \citenamefont {Quintana}, \citenamefont {Roushan}, \citenamefont
  {Vainsencher}, \citenamefont {Wenner}, \citenamefont {Cleland},\ and\
  \citenamefont {Martinis}}]{Kelly2015}%
  \BibitemOpen
  \bibfield  {author} {\bibinfo {author} {\bibfnamefont {J.}~\bibnamefont
  {Kelly}}, \bibinfo {author} {\bibfnamefont {R.}~\bibnamefont {Barends}},
  \bibinfo {author} {\bibfnamefont {A.~G.}\ \bibnamefont {Fowler}}, \bibinfo
  {author} {\bibfnamefont {A.}~\bibnamefont {Megrant}}, \bibinfo {author}
  {\bibfnamefont {E.}~\bibnamefont {Jeffrey}}, \bibinfo {author} {\bibfnamefont
  {T.~C.}\ \bibnamefont {White}}, \bibinfo {author} {\bibfnamefont
  {D.}~\bibnamefont {Sank}}, \bibinfo {author} {\bibfnamefont {J.~Y.}\
  \bibnamefont {Mutus}}, \bibinfo {author} {\bibfnamefont {B.}~\bibnamefont
  {Campbell}}, \bibinfo {author} {\bibfnamefont {Y.}~\bibnamefont {Chen}},
  \bibinfo {author} {\bibfnamefont {Z.}~\bibnamefont {Chen}}, \bibinfo {author}
  {\bibfnamefont {B.}~\bibnamefont {Chiaro}}, \bibinfo {author} {\bibfnamefont
  {A.}~\bibnamefont {Dunsworth}}, \bibinfo {author} {\bibfnamefont {I.-C.}\
  \bibnamefont {Hoi}}, \bibinfo {author} {\bibfnamefont {C.}~\bibnamefont
  {Neill}}, \bibinfo {author} {\bibfnamefont {P.~J.~J.}\ \bibnamefont
  {O/'Malley}}, \bibinfo {author} {\bibfnamefont {C.}~\bibnamefont {Quintana}},
  \bibinfo {author} {\bibfnamefont {P.}~\bibnamefont {Roushan}}, \bibinfo
  {author} {\bibfnamefont {A.}~\bibnamefont {Vainsencher}}, \bibinfo {author}
  {\bibfnamefont {J.}~\bibnamefont {Wenner}}, \bibinfo {author} {\bibfnamefont
  {A.~N.}\ \bibnamefont {Cleland}}, \ and\ \bibinfo {author} {\bibfnamefont
  {J.~M.}\ \bibnamefont {Martinis}},\ }\bibfield  {title} {\emph {\enquote
  {\bibinfo {title} {State preservation by repetitive error detection in a
  superconducting quantum circuit},}\ }}\href {\doibase 10.1038/nature14270}
  {\bibfield  {journal} {\bibinfo  {journal} {Nature}\ }\textbf {\bibinfo
  {volume} {519}},\ \bibinfo {pages} {66} (\bibinfo {year} {2015})}\BibitemShut
  {NoStop}%
\bibitem [{\citenamefont {Schuster}\ \emph {et~al.}(2007)\citenamefont
  {Schuster}, \citenamefont {Houck}, \citenamefont {Schreier}, \citenamefont
  {Wallraff}, \citenamefont {Gambetta}, \citenamefont {Blais}, \citenamefont
  {Frunzio}, \citenamefont {Majer}, \citenamefont {Johnson}, \citenamefont
  {Devoret}, \citenamefont {Girvin},\ and\ \citenamefont
  {Schoelkopf}}]{Schuster2007}%
  \BibitemOpen
  \bibfield  {author} {\bibinfo {author} {\bibfnamefont {D.~I.}\ \bibnamefont
  {Schuster}}, \bibinfo {author} {\bibfnamefont {A.~A.}\ \bibnamefont {Houck}},
  \bibinfo {author} {\bibfnamefont {J.~A.}\ \bibnamefont {Schreier}}, \bibinfo
  {author} {\bibfnamefont {A.}~\bibnamefont {Wallraff}}, \bibinfo {author}
  {\bibfnamefont {J.~M.}\ \bibnamefont {Gambetta}}, \bibinfo {author}
  {\bibfnamefont {A.}~\bibnamefont {Blais}}, \bibinfo {author} {\bibfnamefont
  {L.}~\bibnamefont {Frunzio}}, \bibinfo {author} {\bibfnamefont
  {J.}~\bibnamefont {Majer}}, \bibinfo {author} {\bibfnamefont
  {B.}~\bibnamefont {Johnson}}, \bibinfo {author} {\bibfnamefont {M.~H.}\
  \bibnamefont {Devoret}}, \bibinfo {author} {\bibfnamefont {S.~M.}\
  \bibnamefont {Girvin}}, \ and\ \bibinfo {author} {\bibfnamefont {R.~J.}\
  \bibnamefont {Schoelkopf}},\ }\bibfield  {title} {\emph {\enquote {\bibinfo
  {title} {Resolving photon number states in a superconducting circuit},}\
  }}\href {\doibase 10.1038/nature05461} {\bibfield  {journal} {\bibinfo
  {journal} {Nature}\ }\textbf {\bibinfo {volume} {445}},\ \bibinfo {pages}
  {515} (\bibinfo {year} {2007})}\BibitemShut {NoStop}%
\bibitem [{\citenamefont {Hoffman}\ \emph {et~al.}(2011)\citenamefont
  {Hoffman}, \citenamefont {Srinivasan}, \citenamefont {Schmidt}, \citenamefont
  {Spietz}, \citenamefont {Aumentado}, \citenamefont {T\"ureci},\ and\
  \citenamefont {Houck}}]{Hoffman2011}%
  \BibitemOpen
  \bibfield  {author} {\bibinfo {author} {\bibfnamefont {A.~J.}\ \bibnamefont
  {Hoffman}}, \bibinfo {author} {\bibfnamefont {S.~J.}\ \bibnamefont
  {Srinivasan}}, \bibinfo {author} {\bibfnamefont {S.}~\bibnamefont {Schmidt}},
  \bibinfo {author} {\bibfnamefont {L.}~\bibnamefont {Spietz}}, \bibinfo
  {author} {\bibfnamefont {J.}~\bibnamefont {Aumentado}}, \bibinfo {author}
  {\bibfnamefont {H.~E.}\ \bibnamefont {T\"ureci}}, \ and\ \bibinfo {author}
  {\bibfnamefont {A.~A.}\ \bibnamefont {Houck}},\ }\bibfield  {title} {\emph
  {\enquote {\bibinfo {title} {Dispersive photon blockade in a superconducting
  circuit},}\ }}\href {\doibase 10.1103/PhysRevLett.107.053602} {\bibfield
  {journal} {\bibinfo  {journal} {Phys. Rev. Lett.}\ }\textbf {\bibinfo
  {volume} {107}},\ \bibinfo {pages} {053602} (\bibinfo {year}
  {2011})}\BibitemShut {NoStop}%
\bibitem [{\citenamefont {Kapit}\ \emph {et~al.}(2014)\citenamefont {Kapit},
  \citenamefont {Hafezi},\ and\ \citenamefont {Simon}}]{Kapit2014}%
  \BibitemOpen
  \bibfield  {author} {\bibinfo {author} {\bibfnamefont {E.}~\bibnamefont
  {Kapit}}, \bibinfo {author} {\bibfnamefont {M.}~\bibnamefont {Hafezi}}, \
  and\ \bibinfo {author} {\bibfnamefont {S.~H.}\ \bibnamefont {Simon}},\
  }\bibfield  {title} {\emph {\enquote {\bibinfo {title} {Induced
  self-stabilization in fractional quantum hall states of light},}\ }}\href
  {\doibase 10.1103/PhysRevX.4.031039} {\bibfield  {journal} {\bibinfo
  {journal} {Phys. Rev. X}\ }\textbf {\bibinfo {volume} {4}},\ \bibinfo {pages}
  {031039} (\bibinfo {year} {2014})}\BibitemShut {NoStop}%
\bibitem [{\citenamefont {Houck}\ \emph {et~al.}(2012)\citenamefont {Houck},
  \citenamefont {T\"ureci},\ and\ \citenamefont {Koch}}]{Houck2012}%
  \BibitemOpen
  \bibfield  {author} {\bibinfo {author} {\bibfnamefont {A.~A.}\ \bibnamefont
  {Houck}}, \bibinfo {author} {\bibfnamefont {H.~E.}\ \bibnamefont {T\"ureci}},
  \ and\ \bibinfo {author} {\bibfnamefont {J.}~\bibnamefont {Koch}},\
  }\bibfield  {title} {\emph {\enquote {\bibinfo {title} {On-chip quantum
  simulation with superconducting circuits},}\ }}\href {\doibase
  10.1038/nphys2251} {\bibfield  {journal} {\bibinfo  {journal} {Nat. Phys.}\
  }\textbf {\bibinfo {volume} {8}},\ \bibinfo {pages} {292} (\bibinfo {year}
  {2012})}\BibitemShut {NoStop}%
\bibitem [{\citenamefont {Schmidt}\ and\ \citenamefont
  {Koch}(2013)}]{Schmidt2013}%
  \BibitemOpen
  \bibfield  {author} {\bibinfo {author} {\bibfnamefont {S.}~\bibnamefont
  {Schmidt}}\ and\ \bibinfo {author} {\bibfnamefont {J.}~\bibnamefont {Koch}},\
  }\bibfield  {title} {\emph {\enquote {\bibinfo {title} {Circuit qed lattices:
  Towards quantum simulation with superconducting circuits},}\ }}\href
  {\doibase 10.1002/andp.201200261} {\bibfield  {journal} {\bibinfo  {journal}
  {Annalen der Physik}\ }\textbf {\bibinfo {volume} {525}},\ \bibinfo {pages}
  {395} (\bibinfo {year} {2013})}\BibitemShut {NoStop}%
\bibitem [{\citenamefont {Gambetta}\ \emph {et~al.}(2006)\citenamefont
  {Gambetta}, \citenamefont {Blais}, \citenamefont {Schuster}, \citenamefont
  {Wallraff}, \citenamefont {Frunzio}, \citenamefont {Majer}, \citenamefont
  {Devoret}, \citenamefont {Girvin},\ and\ \citenamefont
  {Schoelkopf}}]{Gambetta2006}%
  \BibitemOpen
  \bibfield  {author} {\bibinfo {author} {\bibfnamefont {J.}~\bibnamefont
  {Gambetta}}, \bibinfo {author} {\bibfnamefont {A.}~\bibnamefont {Blais}},
  \bibinfo {author} {\bibfnamefont {D.~I.}\ \bibnamefont {Schuster}}, \bibinfo
  {author} {\bibfnamefont {A.}~\bibnamefont {Wallraff}}, \bibinfo {author}
  {\bibfnamefont {L.}~\bibnamefont {Frunzio}}, \bibinfo {author} {\bibfnamefont
  {J.}~\bibnamefont {Majer}}, \bibinfo {author} {\bibfnamefont {M.~H.}\
  \bibnamefont {Devoret}}, \bibinfo {author} {\bibfnamefont {S.~M.}\
  \bibnamefont {Girvin}}, \ and\ \bibinfo {author} {\bibfnamefont {R.~J.}\
  \bibnamefont {Schoelkopf}},\ }\bibfield  {title} {\emph {\enquote {\bibinfo
  {title} {Qubit-photon interactions in a cavity: Measurement-induced dephasing
  and number splitting},}\ }}\href {\doibase 10.1103/PhysRevA.74.042318}
  {\bibfield  {journal} {\bibinfo  {journal} {Phys. Rev. A}\ }\textbf {\bibinfo
  {volume} {74}},\ \bibinfo {pages} {042318} (\bibinfo {year}
  {2006})}\BibitemShut {NoStop}%
\bibitem [{\citenamefont {Clerk}\ \emph {et~al.}(2010)\citenamefont {Clerk},
  \citenamefont {Devoret}, \citenamefont {Girvin}, \citenamefont {Marquardt},\
  and\ \citenamefont {Schoelkopf}}]{Clerk2010}%
  \BibitemOpen
  \bibfield  {author} {\bibinfo {author} {\bibfnamefont {A.~A.}\ \bibnamefont
  {Clerk}}, \bibinfo {author} {\bibfnamefont {M.~H.}\ \bibnamefont {Devoret}},
  \bibinfo {author} {\bibfnamefont {S.~M.}\ \bibnamefont {Girvin}}, \bibinfo
  {author} {\bibfnamefont {F.}~\bibnamefont {Marquardt}}, \ and\ \bibinfo
  {author} {\bibfnamefont {R.~J.}\ \bibnamefont {Schoelkopf}},\ }\bibfield
  {title} {\emph {\enquote {\bibinfo {title} {Introduction to quantum noise,
  measurement, and amplification},}\ }}\href {\doibase
  10.1103/RevModPhys.82.1155} {\bibfield  {journal} {\bibinfo  {journal} {Rev.
  Mod. Phys.}\ }\textbf {\bibinfo {volume} {82}},\ \bibinfo {pages} {1155}
  (\bibinfo {year} {2010})}\BibitemShut {NoStop}%
\bibitem [{Note9()}]{Note9}%
  \BibitemOpen
  \bibinfo {note} {We define potential trenches as domains of connected
  plaquettes identified by $U > T$.}\BibitemShut {Stop}%
\bibitem [{Note10()}]{Note10}%
  \BibitemOpen
  \bibinfo {note} {Our 1D model readily applies, for any temperature $T$, to
  implementations of our scheme based on 1D stabilizer codes (see, e.g.,
  Ref.~\cite {Pedrocchi2015}).}\BibitemShut {Stop}%
\bibitem [{Note11()}]{Note11}%
  \BibitemOpen
  \bibinfo {note} {The first term corresponds to the typical time required for
  anyons to diffuse and recombine. The second corresponds to the subsequent
  trench decay.}\BibitemShut {Stop}%
\bibitem [{\citenamefont {Gillespie}(1977)}]{Gillespie1977}%
  \BibitemOpen
  \bibfield  {author} {\bibinfo {author} {\bibfnamefont {D.~T.}\ \bibnamefont
  {Gillespie}},\ }\bibfield  {title} {\emph {\enquote {\bibinfo {title} {Exact
  stochastic simulation of coupled chemical reactions},}\ }}\href {\doibase
  10.1021/j100540a008} {\bibfield  {journal} {\bibinfo  {journal} {J. Phys.
  Chem.}\ }\textbf {\bibinfo {volume} {81}},\ \bibinfo {pages} {2340} (\bibinfo
  {year} {1977})}\BibitemShut {NoStop}%
\bibitem [{\citenamefont {Jansen}(1995)}]{Jansen1995}%
  \BibitemOpen
  \bibfield  {author} {\bibinfo {author} {\bibfnamefont {A.~P.~J.}\
  \bibnamefont {Jansen}},\ }\bibfield  {title} {\emph {\enquote {\bibinfo
  {title} {Monte carlo simulations of chemical reactions on a surface with
  time-dependent reaction-rate constants},}\ }}\href {\doibase
  10.1016/0010-4655(94)00155-U} {\bibfield  {journal} {\bibinfo  {journal}
  {Comp. Phys. Comm.}\ }\textbf {\bibinfo {volume} {86}},\ \bibinfo {pages} {1}
  (\bibinfo {year} {1995})}\BibitemShut {NoStop}%
\bibitem [{\citenamefont {Edmonds}(1965)}]{Edmonds1965}%
  \BibitemOpen
  \bibfield  {author} {\bibinfo {author} {\bibfnamefont {J.}~\bibnamefont
  {Edmonds}},\ }\bibfield  {title} {\emph {\enquote {\bibinfo {title} {Paths,
  trees, and flowers},}\ }}\href {\doibase 10.4153/CJM-1965-045-4} {\bibfield
  {journal} {\bibinfo  {journal} {Canad. J. Math.}\ }\textbf {\bibinfo {volume}
  {17}},\ \bibinfo {pages} {449} (\bibinfo {year} {1965})}\BibitemShut
  {NoStop}%
\bibitem [{\citenamefont {Kolmogorov}(2009)}]{Kolmogorov2009}%
  \BibitemOpen
  \bibfield  {author} {\bibinfo {author} {\bibfnamefont {V.}~\bibnamefont
  {Kolmogorov}},\ }\bibfield  {title} {\emph {\enquote {\bibinfo {title}
  {Blossom v: a new implementation of a minimum cost perfect matching
  algorithm},}\ }}\href {\doibase 10.1007/s12532-009-0002-8} {\bibfield
  {journal} {\bibinfo  {journal} {Math. Prog. Comp.}\ }\textbf {\bibinfo
  {volume} {1}},\ \bibinfo {pages} {43} (\bibinfo {year} {2009})}\BibitemShut
  {NoStop}%
\bibitem [{\citenamefont {Pedrocchi}\ \emph {et~al.}(2015)\citenamefont
  {Pedrocchi}, \citenamefont {Bonesteel},\ and\ \citenamefont
  {DiVincenzo}}]{Pedrocchi2015}%
  \BibitemOpen
  \bibfield  {author} {\bibinfo {author} {\bibfnamefont {F.~L.}\ \bibnamefont
  {Pedrocchi}}, \bibinfo {author} {\bibfnamefont {N.~E.}\ \bibnamefont
  {Bonesteel}}, \ and\ \bibinfo {author} {\bibfnamefont {D.~P.}\ \bibnamefont
  {DiVincenzo}},\ }\bibfield  {title} {\emph {\enquote {\bibinfo {title} {Monte
  carlo studies of the self-correcting properties of the majorana quantum error
  correction code under braiding},}\ }}\href {\doibase
  10.1103/PhysRevB.92.115441} {\bibfield  {journal} {\bibinfo  {journal} {Phys.
  Rev. B}\ }\textbf {\bibinfo {volume} {92}},\ \bibinfo {pages} {115441}
  (\bibinfo {year} {2015})}\BibitemShut {NoStop}%
\end{thebibliography}%

\end{document}